\documentclass[a4paper,11pt]{grl}

\usepackage[T1]{fontenc}
\usepackage{url}
\usepackage{natbib}
\usepackage{amsmath}
\usepackage{textcomp}
\usepackage{gensymb}
\usepackage{graphicx}
\usepackage{enumitem}
\usepackage[utf8]{inputenc}
\usepackage{setspace}

\DeclareRobustCommand{\VAN}[3]{#2}

\journalname{Geophysical Research Letters}

\begin{document}

\title{Evidence for a kilometre-scale seismically slow layer atop the core-mantle boundary from normal modes}

\authors{Stuart Russell\affil{1,2}, Jessica C.E. Irving\affil{3}, Lisanne Jagt\affil{1}, Sanne Cottaar\affil{1}}

\affiliation{1}{University of Cambridge}
\affiliation{2}{Universit\"at M\"unster}
\affiliation{3}{University of Bristol}

\correspondingauthor{Stuart Russell}{srussell@uni-muenster.de}

\begin{keypoints}

\item Normal mode centre frequencies are a sensitive data type to detect seismically anomalous thin layers at the core-mantle boundary.

\item A slow and dense layer on the order of 1~km thick atop the core-mantle boundary can improve the fit to normal mode data.

\item The inclusion of a layer is likely not a unique way to improve the 1D model and layer properties remain uncertain.

\end{keypoints}

\section*{Plain language summary}

Normal modes are long-period oscillations of the whole Earth as it vibrates after large earthquakes. The frequency that a mode oscillates at depends on the interior structure of the Earth. Research suggests a global and thin layer of anomalous composition and low seismic wave speeds may have formed at the base of Earth’s mantle, but would be difficult to observe seismically. We test and quantify the effect of this layer on the frequencies at which normal modes vibrate. We then compare these predictions to a large dataset of normal mode frequency measurements to examine whether such a layer is consistent with observed data. We find that not only is a layer of 1~-~3~km thickness permitted by the modes, but that a layer being present improves the fit to the data. There are a wide range of parameters that adequately fit the dataset so we cannot be specific about its properties. Furthermore a layer is likely not a unique way to improve the model. A seismically slow layer at the core-mantle boundary has implications for processes in the mantle and outer core and the interaction between them.

\begin{abstract}

Geodynamic modelling and seismic studies have highlighted the possibility that a thin layer of low seismic velocities, potentially molten, may sit atop the core-mantle boundary but has thus far eluded detection. In this study we employ normal modes, an independent data type to body waves, to assess the visibility of a seismically slow layer atop the core-mantle boundary to normal mode centre frequencies. Using forward modelling and a dataset of 353 normal mode observations we find that some centre frequencies are sensitive to one-dimensional kilometre-scale structure at the core-mantle boundary. Furthermore, a global slow and dense layer 1~-~3~km thick is better-fitting than no layer. The well-fitting parameter space is broad with a wide range of possible seismic parameters, which precludes inferring a possible composition or phase. Our methodology cannot uniquely detect a layer in the Earth but one should be considered possible and accounted for in future studies.

\end{abstract}

\section{Introduction}

The Earth's core-mantle boundary (CMB) is a major internal boundary in the planet, separating the liquid iron outer core from the solid silicate mantle. Structures and processes at the CMB affect convection in both the mantle and the outer core, with consequences for plate tectonics and the geomagnetic field. The lowermost mantle immediately atop the CMB is extremely complex; it contains numerous structures identified from seismology including ultra-low velocity zones (ULVZs). 

ULVZs are extreme seismic anomalies tens of kilometres thick and hundreds of kilometres in diameter that exhibit velocity reductions on the order of 10\% and 30\% for P and S waves, respectively \citep[e.g.][]{Yu2018}. There are several possible origins of ULVZs including iron-enrichment \citep[e.g.][]{Wicks2010, Dobrosavljevic2019}, the presence of melt or partial melt \citep[e.g.][]{Williams1996, Yuan2017}, or both combined \citep{Liu2016, Dannberg2021}. Melt of any origin in the lowermost mantle is expected to be dense \citep{Boukare2015} and should drain from the ULVZ under gravity \citep{Hernlund2007, Dannberg2021}. If drainage occurs then ULVZs are expected to be accompanied by an extensive fully-molten layer which may have important implications for geochemical observations \citep[e.g.][]{Coltice2011} and for electromagnetic coupling between the core and mantle \citep[e.g.][]{Buffett2002, Holme2013}.

Body waves are commonly used to study ULVZs, including reflected phases \citep[e.g.][]{Brown2015}, transmitted phases \citep[e.g.][]{Zou2007} and diffracted phases \citep[e.g.][]{Rost2006, Li2022}. Despite body waves being sensitive to ULVZs, they may not be sensitive to an accompanying layer if it is too thin. Travel times of PKKP\textsubscript{diff} relative to PKKP\textsubscript{bc} are sensitive to and consistent with a thin layer on the order of kilometres thick, but show significant scatter impeding a conclusion on whether a layer exists or not \citep{Russell2022}.

In contrast to body waves, normal modes are very long-period free oscillations of the Earth excited after large earthquakes (M\textsubscript{w} $\ge$\ 7.4). Normal modes appear as prominent peaks in the Fourier spectra of several days-long seismograms. The shape and frequencies of these peaks depend on the internal structure of the planet \citep[e.g.][]{Dziewonski1971, Dziewonski1981}. The degenerate frequencies that the mode would oscillate at in the absence of 3D structure, called centre frequencies, can be estimated from seismic spectra and give information about average 1D Earth structure. Constraining lateral heterogeneity requires study of spectra or the higher order coefficients of mode splitting functions \citep{Woodhouse1986}. Stoneley modes have peak sensitivity at the CMB interface \citep{Stoneley1924, Koelemeijer2013}. Splitting functions can observe small-scale ULVZs if there is a long-wavelength component in their lateral distribution \citep{Koelemeijer2012}.

Building upon the results of \cite{Russell2022}, we use forward modelling to test the sensitivity of normal mode centre frequencies to a kilometre-scale global low velocity layer atop the CMB. We then compare the modelling results to a dataset of centre frequency measurements to ascertain whether a globally layered CMB is compatible with seismic observations.

\section{Data} \label{data}

Measuring centre frequencies of normal modes is non-trivial and the centre frequencies and splitting functions are calculated from spectra by inversion. We do not measure centre frequencies ourselves, but instead use published datasets. For this study we use the compilation used to construct the Elastic Parameters of the Outer Core (EPOC) model \citep{Irving2018}, together with Stoneley modes \citep{Koelemeijer2013}, the longest period modes \citep{Deuss2011}, and recently measured toroidal modes \citep{Schneider2021}. This excludes inner-core sensitive modes, which may couple strongly due to unmodelled inner core anisotropy \citep{Irving2008}. Centre frequencies and measurement errors are taken from the most recent publication in which that mode is measured \citep{Masters1995, Resovsky1998, Widmer2002, Deuss2013, Koelemeijer2013, Schneider2021, REM}, giving a total of 353 modes. Centre frequencies are corrected for the first-order effects of Earth's ellipticity \citep{Dahlen1999}. A table of centre frequencies and errors can be found in Section~S11.

\section{Normal mode sensitivity to a layer} \label{further}

We use MINEOS \citep{Woodhouse1988, Mineos} to forward model the eigenfrequencies (centre frequencies) and eigenfunctions of the free oscillations of a given model. Slow and dense layers of different thicknesses are inserted above the CMB into a 1D model. This 1D model, referred to as EPOC, consists of PREM \citep{Dziewonski1981} for the mantle and inner core, while EPOC is used for the outer core as it improves the misfit for outer core sensitive modes.

We first investigate the sensitivity of the modes to the presence of a layer. The parameters for one test layer are $-$20\%, $-$40\% and $+$50\% for $\delta V_{p}$, $\delta V_{s}$ and $\delta \rho$ relative to PREM, respectively (absolute: 10.97~$\mathrm{km \, s^{-1}}$, 4.36~$\mathrm{km \, s^{-1}}$, 8.35~$\mathrm{g \, cm^{-3}}$). As it has non-zero $V_{s}$ we refer to this layer as the `solid layer', however it could represent partial melt. We test a second layer with $-$33\%, $-$100\% and $+$14\% for $\delta V_{p}$, $\delta V_{s}$ and $\delta \rho$, respectively (absolute: 9.10~$\mathrm{km \, s^{-1}}$, 0.00~$\mathrm{km \, s^{-1}}$, 6.35~$\mathrm{g \, cm^{-3}}$), and refer to this as the `molten layer'. This layer is identical to that explored by \cite{Russell2022} using PKKP diffracted waves.

\begin{figure}
    \centering
    \includegraphics[width=14cm]{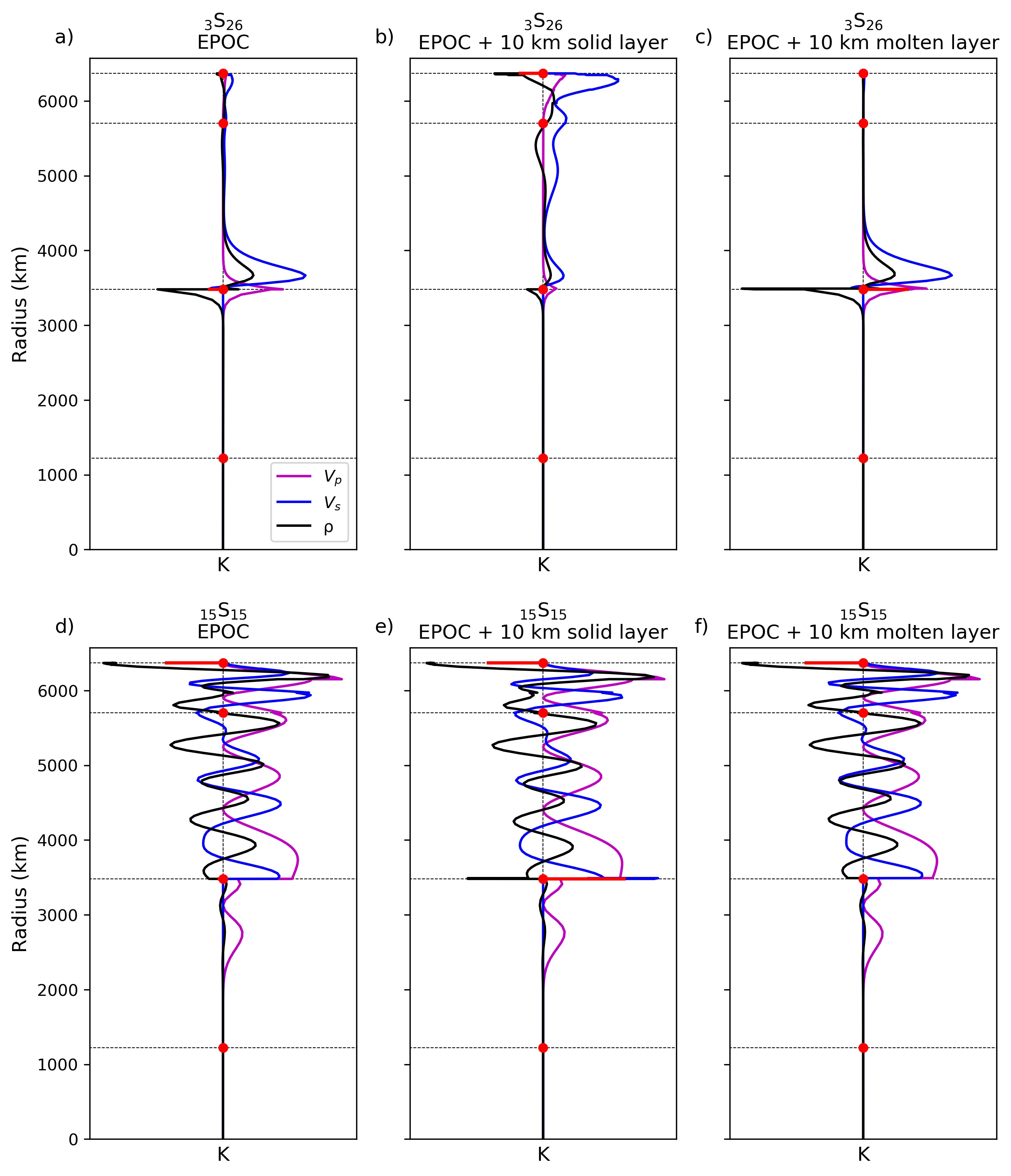}
    \caption{Sensitivity kernels, $K$, for two different modes in three different models indicated by subfigure titles, showing sensitivity to P-velocity (magenta), S-velocity (blue) and density (black). Horizontal dashed lines represent the major discontinuities: the surface, 660~km, CMB and ICB; and discontinuity sensitivity to displacement is given by red horizontal lines. Amplitudes are consistent across plots of the same mode.}
    \label{kernels_main}
\end{figure}

Using the eigenfunctions from MINEOS we calculate sensitivity kernels \citep{Dahlen1999}. Kernels are used to examine the sensitivity of a mode and to assess how sensitivity evolves as the model is varied. Figure~\ref{kernels_main} shows sensitivity kernels for \textsubscript{3}S\textsubscript{26} and \textsubscript{15}S\textsubscript{15} in the cases of EPOC, EPOC with a 10~km solid layer, and EPOC with a 10~km molten layer. For both modes, the sensitivity evolves when layers are included. \textsubscript{3}S\textsubscript{26} shows a drastic change in character; it is a Stoneley mode in EPOC and for a molten layer, but not for a solid layer where the CMB sensitivity is reduced. The kernel for \textsubscript{15}S\textsubscript{15} does not significantly change between different models, but does acquire a peak inside the solid layer. That sensitivity peaks in such thin regions validates the use of normal modes to study thin structures at the CMB, in contrast to their efficacy when investigating less extreme structures \citep[e.g.][]{Robson2022}. Although we illustrate these extreme examples, the majority of kernels do not notably change. Further sensitivity kernels are presented in Section~S1.

Sensitivity kernels for the solid and molten layers often differ and this corresponds to drastic differences in behaviour of the eigenfrequencies as layer thickness increases (Section~S2). The model-dependent sensitivity kernels means that a linearised Backus-Gilbert style inversion \citep{Backus1970, Masters2003} cannot be applied to the problem of resolving extreme layers at the CMB.

\begin{figure}
    \centering
    \includegraphics[width=14cm]{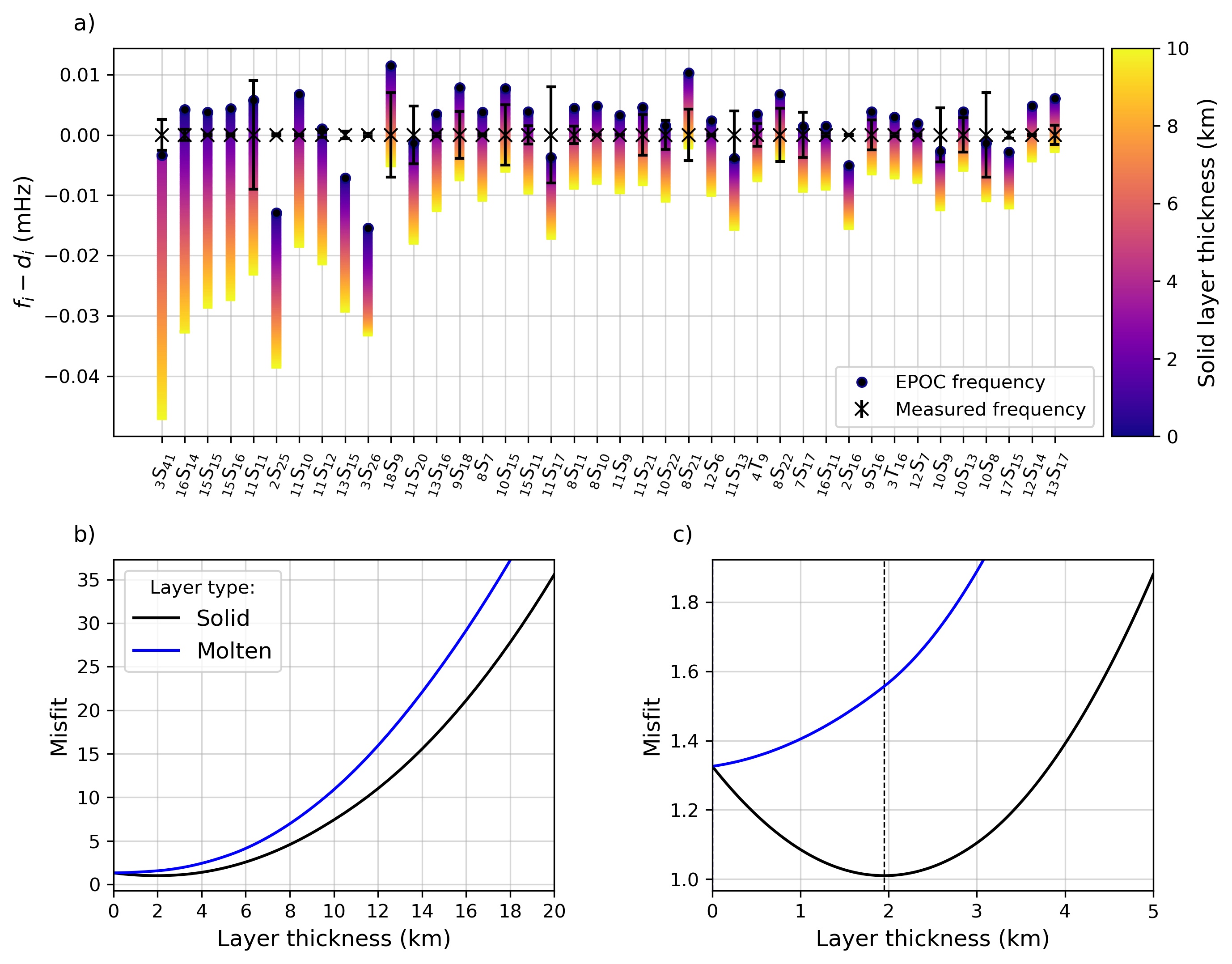}
    \caption{(a) Colours show eigenfrequencies predicted for different thicknesses of the solid layer, $f_{i}$, relative to the measured centre frequencies, $d_{i}$. Measurement errors, $\epsilon_{d}$, are the black error bars. These are the 40 most sensitive modes ordered by the gradient of their eigenfrequencies with solid layer thickness. A plot showing (a) for all 353 modes used to calculate $M$ can be found in Section~S3. (b) and (c) show the misfit, $M$, as a function of layer thickness between 0~-~20~km and 0~-~5~km,  respectively, for the solid ($\delta V_{p}$:$-$20\%, $\delta V_{s}$:$-$40\% and $\delta \rho$:$+$50\%) and molten ($\delta V_{p}$:$-$33\%, $\delta V_{s}$:$-$100\% and $\delta \rho$:$+$14\%) layers. The vertical dotted line marks the minimum in $M$ for the solid layer.}
    \label{misfit_solid}
\end{figure}

\section{Fit to Mode Dataset}

For each model, the best-fitting layer thickness is assessed by calculating a least-squares misfit, $M$, between the observed, $d$, and predicted, $f$, centre frequencies using $N=353$ modes. The misfit is weighted by the errors; in addition to the published `measurement error', $\epsilon_d$, we also associate a `model error', $\epsilon_f$, which is the range of eigenfrequency values from PREM, STW105 \citep{Kustowski2008} and EPOC, representing the variation of a mode between different 1D models. All error values used are given in Section~S11.

\begin{equation}
    M = \frac{1}{N} \sum_{i=0}^{N} \frac{\left( f_{i} - d_{i} \right)^{2}}{ \epsilon_{f_{i}}^{2} + \epsilon_{d_{i}}^{2}}.
\end{equation}

The minimum misfit and best-fitting layer thickness are calculated for both layers (Figure~\ref{misfit_solid}). For many modes, the presence of a solid layer improves the fit to the measurement (Figure~\ref{misfit_solid}a). For the solid layer, the best-fitting layer thickness is 1.95~km, but for the molten layer is 0~km (Figures~\ref{misfit_solid}b-c), indicating that the modes do not prefer a molten layer and the upwards movement of the solid-fluid interface over EPOC.

We also test other background models. PREM and STW105 have best-fitting solid layer thicknesses of 2.7~km and 2.75~km, respectively (Section~S4). These models differ in the upper mantle, while PREM and EPOC differ in the outer core. The different results for PREM and EPOC indicate a trade-off between structure above and below the CMB \citep[see also][]{Irving2018, vanTent2020}. Additionally, we show that toroidal modes independently prefer a layer (Section~S5). There is also a trade-off between the strength of the seismic anomalies and the layer thicknesses (Section~S6), and with lower mantle velocity structure and anisotropy (Section~S7). We perform simple tests of the effects of anisotropy by including an anisotropic D\textsuperscript{$\prime \prime$} \citep[as in][]{Montagner1996} and anisotropy throughout the lower mantle \citep[following][]{deWit2015}, neither of which affects the preferred layer thickness by more than 150~m. The misfit reduction cannot be achieved by moving the CMB alone (Section~S8). We also test the robustness of our results by perturbing the measured and predicted centre frequencies according to the errors (Section~S9); the data prefer a layer on the order of 1~km thick even accounting for underestimation in the measured uncertainties \citep[e.g.][]{Akbarashrafi2018}.

\section{Grid Search for Layer Properties} \label{inital}

\begin{figure}
    \centering
    \includegraphics[width=14cm]{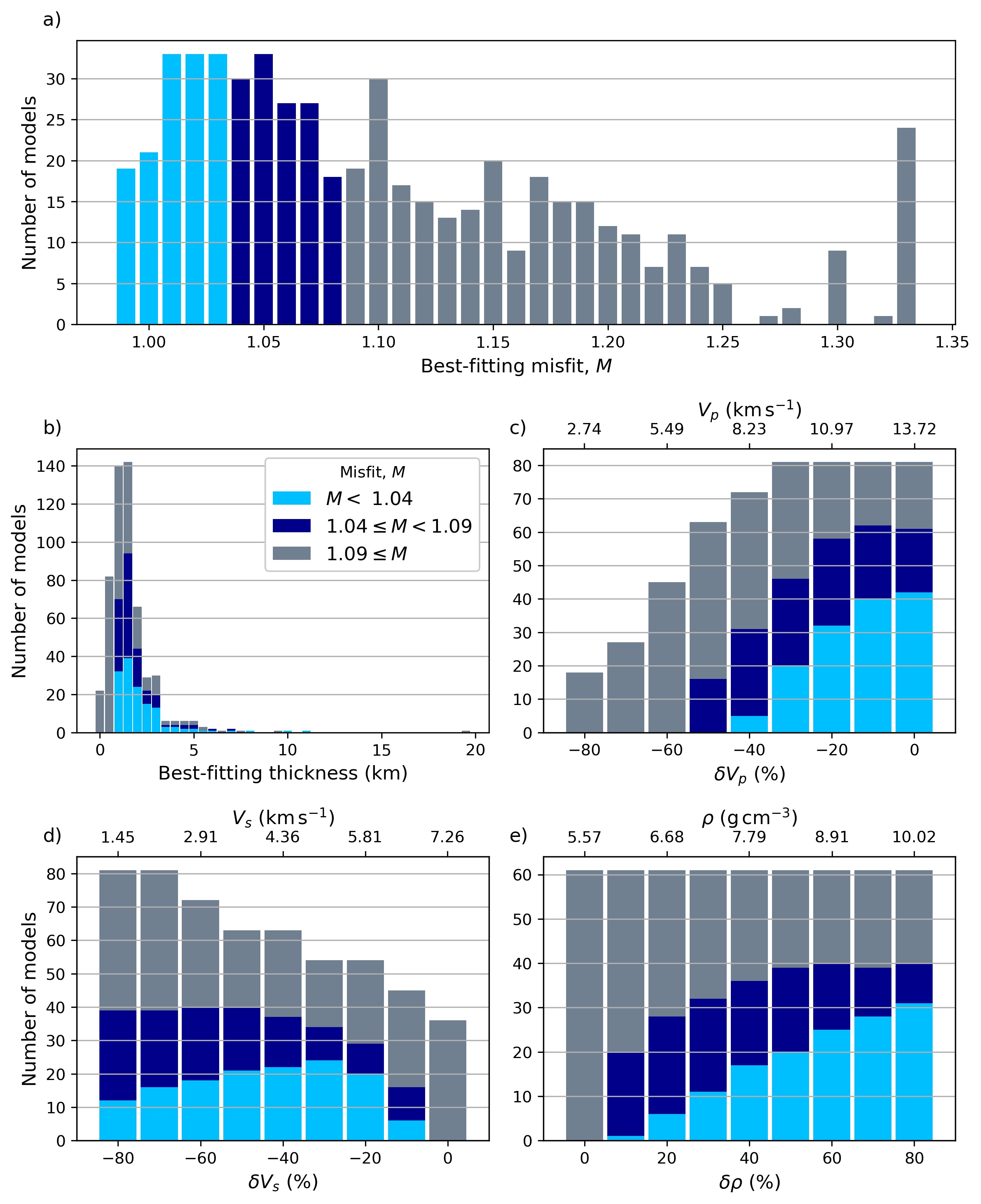}
    \caption{(a) The distribution of the best-fitting misfit values. Colours correspond to the misfit. Models in light blue ($M$\textless 1.04) are referred to as `well-fitting'. (b) Best-fitting layer thicknesses coloured by misfit. (c-e) Distribution of misfit values across the range of $\delta V_{p}$,  $\delta V_{s}$ and $\delta \rho$.}
    \label{grid search}
\end{figure}

In order to further constrain the parameters, we construct a grid search over a wide parameter space, excluding a fully molten scenario. $\delta V_{p}$ and $\delta V_{s}$ were varied from $-$80\% to 0\% relative to PREM, while $\delta \rho$ was varied from 0\% to $+$80\%. All properties were varied in intervals of 10\%. We exclude models which have negative bulk modulus and all 549 remaining combinations of parameters were tested for layer thicknesses from 0~km to 20~km in 0.5~km intervals.

Figure~\ref{grid search} summarises the results of this search (in-depth results in Section~S10). EPOC with no layer has a misfit of 1.33 and the best-fitting misfit found in the search was 0.99 (Figure~\ref{grid search}a), an improvement of 26\%. 139 models have misfits within 0.05 of this global minimum, which we term `well-fitting'; our initially-tested solid layer meets this criterion. The best-fitting thicknesses are concentrated at values less than 3~km indicating that the modes are generally better-fitted by thinner layers (Figure~\ref{grid search}b). All models with extreme (> 40\%) values of $\delta V_{p}$ fit poorly (Figure~\ref{grid search}c). In contrast, there is a wide range of well-fitting values of $\delta V_{s}$ ($-$10\% to $-$80\%, Figure~\ref{grid search}d) and for density ($+$10\% to $+$80\%, Figure~\ref{grid search}e). It is notable that 96\% of the models tested preferred the inclusion of a layer, however the wide parameter space precludes further constraint.

\section{Discussion}

\subsection{Limitations}

Our methodology does not allow us to verify that a layer must exist, but illustrates an improvement to EPOC that reduces the misfit to observations. Our results indicate that a thin layer at the CMB can significantly improve the fit to normal modes. If such a structure exists, which is reasonable to suspect, then it may have not arisen in existing global 1D models due to their smooth parameterisation. Future models should therefore consider this layer in their construction. It is well known that the lowermost mantle is strongly heterogeneous and whether our results can be (partly) explained by the averaged signal of heterogeneity, such as ULVZs, and/or CMB topography requires further testing.

\subsection{Comparison to Body Wave Data}

PKKP\textsubscript{diff} and PKKP\textsubscript{bc} differential times are sensitive to the presence of thin layers atop the CMB \citep{Russell2022}. Using \textgreater 12,500 observations, \cite{Russell2022} concluded a seismically anomalous layer $\sim$1~-~2~km thick was consistent with their data, but that robust observation was inhibited by scatter. 3D structures such as ULVZs \citep{Chen2023} and CMB topography may be a significant cause of this scatter. 

Normal modes centre frequencies do not have sensitivity to these 3D structures. Furthermore, a totally molten layer in EPOC is not preferred by normal modes. This does not contradict the results of \cite{Russell2022} as PKKP\textsubscript{diff} phases are only sensitive to $V_{p}$, while normal modes are also sensitive to $V_{s}$ and $\rho$. The 1~-~2~km of $-$33\% $V_{p}$ that fits the PKKP\textsubscript{diff} observations is consistent with the results of this study. Our finding that numerous models with smaller $\delta V_{s}$ are well-fitting indicates that an examination of less extreme layers with body waves, especially CMB sensitive S-waves such as ScP, S\textsubscript{diff} or SPdKS \citep[e.g.][]{Brown2015, Thorne2021, Li2022}, may be warranted. Recent PcP observations have been used to infer the existence of a global layer of subducted material \citep{Hansen2023} and S\textsubscript{diff} observations have been found to be consistent with a slow layer underlying the entire Pacific \citep[][]{Wolf2023}.

Other studies have highlighted that ULVZs may be internally layered with higher proportions of melt, iron or both at their base \citep{Thybo2003, Ross2004, Li2022}. Such a layer would be seismically more anomalous and, if it was laterally more extensive than the ULVZs themselves, may be the structure that the normal modes are sensing. \cite{Ross2004} observed a ULVZ beneath Siberia and found that PcP data are fitted by a 6~km thick lower layer with $-$25\% $\delta V_{p}$ and $-$45\% $\delta V_{s}$. These parameters agree with our constraints, and the discrepancy in thickness could be due to their observations being within a ULVZ, while ours represent a global average. Additionally \cite{Li2022} find that S\textsubscript{diff} post-cursors through the Hawaiian ULVZ suggest a 2~km thick lower basal layer with $-$40\% $\delta V_{s}$.

\subsection{Layer Origins and Implications}

A low velocity layer at at the CMB may originate from melting \citep{Williams1996, Dannberg2021} or iron enrichment \citep{Wicks2010, Dobrosavljevic2019}. It could be the partially-molten or solidified remnants of a basal magma ocean \citep{Labrosse2007}, the product of core-mantle chemical exchange \citep[e.g.][]{Stein2023}, buoyant core-derived solids \citep{mittal2020, Fu2023}, or subducted material \citep[e.g.][]{Hansen2023}. A partially molten lowermost mantle has been proposed for both the moon \citep{Weber2011} and Mars \citep{Samuel2021}.

\begin{figure}
    \centering
    \includegraphics[width=14cm]{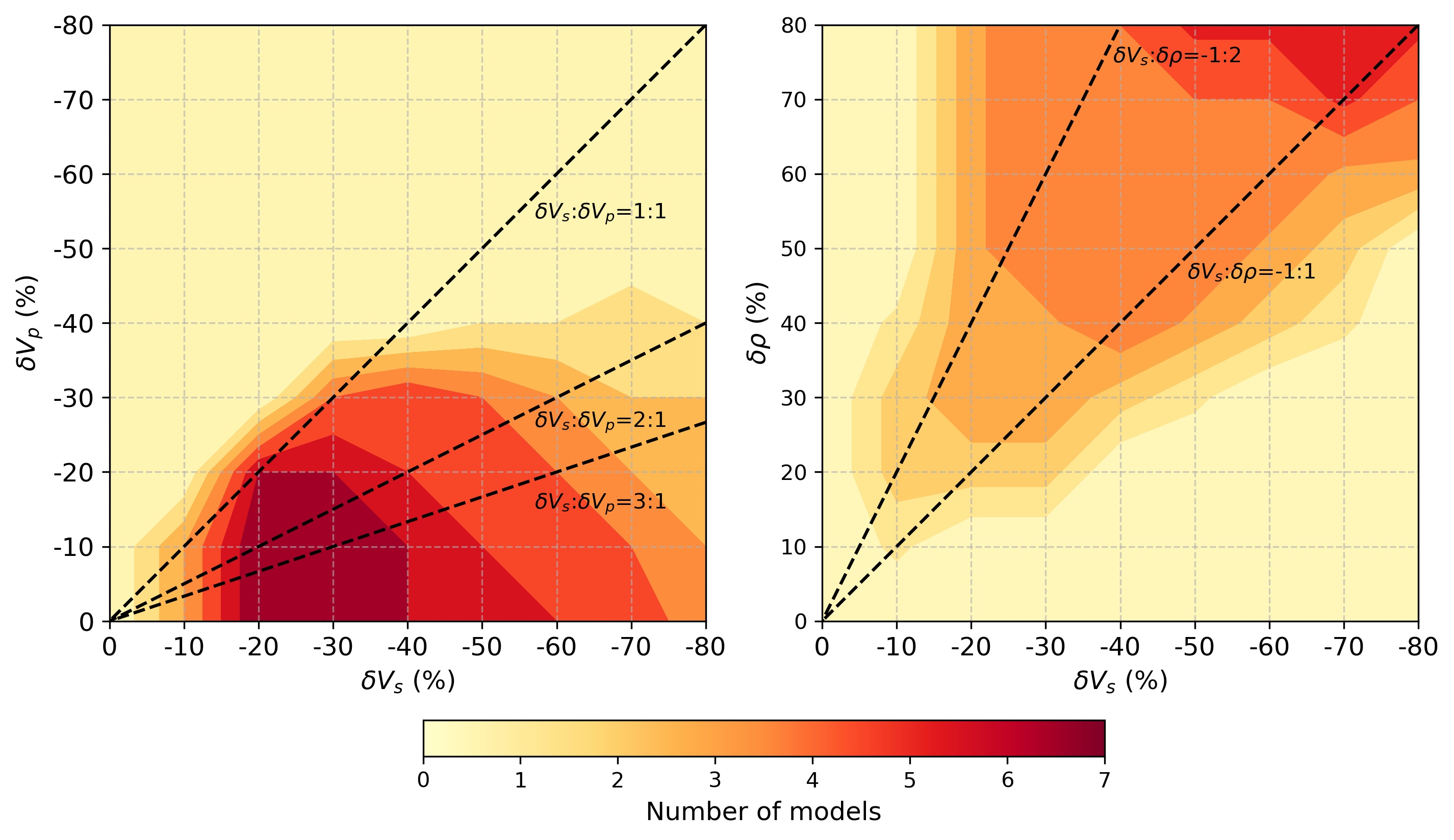}
    \caption{(a) The density of well-fitting models as a function of $\delta V_{s}$ and $\delta V_{p}$. (b) as for (a) but for $\delta V_{s}$ and $\delta \rho$. Dashed lines show anomaly ratios as labelled.}
    \label{ratios}
\end{figure}

\begin{figure}
    \centering
    \includegraphics[width=14cm]{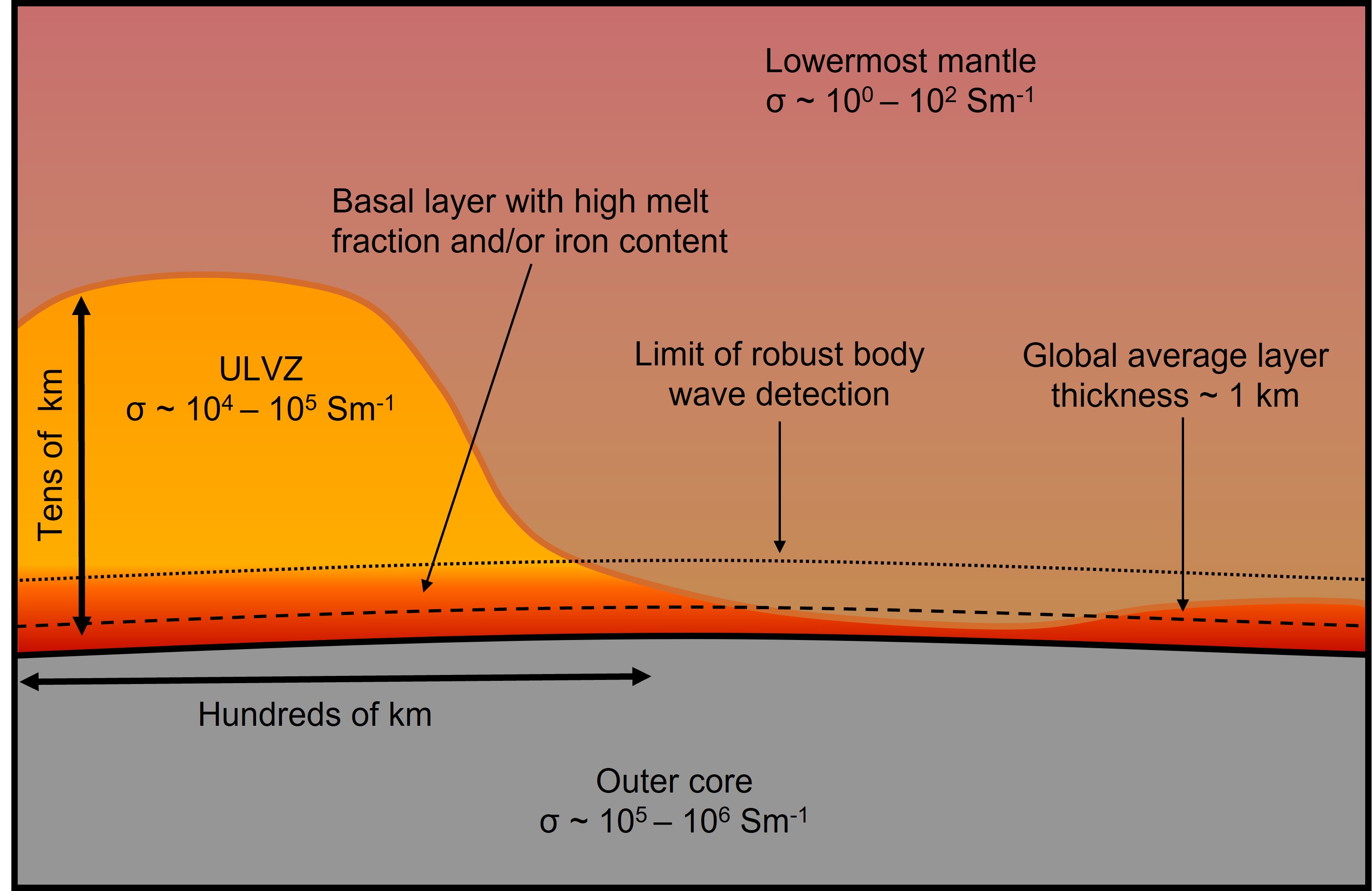}
    \caption{Cartoon demonstrating how a layer may relate to ULVZs and core-mantle interaction. $\sigma$ indicates electrical conductivity.}
    \label{cartoon}
\end{figure}

Figure~\ref{ratios} shows the occurrence of well-fitting models across the parameter space. Most well-fitting models have a $\delta V_{s}$:$\delta V_{p}$ ratio exceeding 1:1 and many exceed 3:1. Iron-rich ferropericlase \citep{Dobrosavljevic2019} and partial melt \citep{Berryman2000, Dannberg2021} can explain these ratios. Dense partial melt in the lowermost mantle should percolate downwards, resulting in a totally molten layer atop the CMB \citep{Hernlund2007, Dannberg2021} which is not favoured by the modes. The decrease in $\delta V_{s}$ and increase in $\delta \rho$ are relatively well-correlated in the well-fitting models which may support an iron-rich composition. Due to the wide range of well-fitting parameters we do not infer the layer's origins, however the strongly favoured density increase may suggest a structure that is distinct from currently described ULVZs, which typically have inferred density increases up to 20\% \citep{Yu2018}. 

Figure~\ref{cartoon} demonstrates how a global low velocity layer may relate to ULVZs and other CMB phenomena. Constraints on lowermost mantle electrical conductivity come from observations of Earth's length-of-day and diurnal nutations, which require electromagnetic coupling between the core and mantle with a total conductance of 10\textsuperscript{8}~S \citep{Buffett2002, Holme1998}. The electrical conductivity of silicates in the lowermost mantle is on the order of 10\textsuperscript{0}~-~10\textsuperscript{2}~Sm\textsuperscript{-1} \citep{Sinmyo2014}, however for post-perovskite it could be greater \citep{Ohta2008}. Coupling is limited to the skin depth of magnetic diffusion, which for diurnal nutations is $\sim$500~m, requiring a lowermost mantle conductivity on the order of 10\textsuperscript{5}~Sm\textsuperscript{-1} \citep{Buffett2002}. The conductivity of ferropericlase at CMB conditions is on the order of 10\textsuperscript{4}~-~10\textsuperscript{5}~Sm\textsuperscript{-1} \citep{Holmstrom2018, Ho2023}, and thus iron-enrichment can satisfy both the conductance requirement and the seismology.

This layer may also provide an anomalous geochemical reservoir. Melt or iron at the CMB may enhance core-mantle interaction leading to geochemical anomalies detected at hotspots \citep{Brandon1998, Mundl2020}, or could itself be a primordial reservoir \citep{Williams2015}. Disentangling geochemical signatures that may relate to a layer of unknown origin would be difficult, however our results suggest that this should be considered.

\section{Conclusions}

We have used a dataset of 353 normal mode centre frequency measurements, coupled with a forward modelling approach, to test the visibility of a global kilometre-scale slow layer atop the CMB to normal mode centre frequencies. We find the fit of EPOC to the dataset is improved by a global slow and dense layer with an average thickness on the order of one kilometre. However, a wide range of well-fitting seismic parameters, trade-offs with layer thickness and with the choice of background model make assessment of the exact layer properties and thickness impossible.

Broadly speaking, the best-fitting models are 1~-~3~km thick, have a tens of percent (10-80\%) decrease in S-velocity and increase in density, and a lesser decrease (0-40\%) in P-velocity relative to PREM's lowermost mantle. There is a preference for high $\delta V_{s}$:$\delta V_{p}$ ratios, which could be indicative of either partial melt or iron enrichment at the CMB. Forward modelling of normal modes cannot prove this layer is required but, when constructing future models of the Earth, it should be considered a viable structure that may improve the fit to seismic data. If a layer exists at the CMB, it would have multi-disciplinary implications for core-mantle coupling, geochemistry, and Earth's evolution.

\acknowledgments

We thank members of the University of Cambridge Earth Science Department, specifically Carl Martin, Alex Myhill, Pallav Kant, Adam Butler, David Al-Attar and Helen Williams, as well as Tim Elliot, Simon Lock, Robert Myhill, Sebastian Rost and R\^{u}na van Tent for insightful scientific discussions. Thanks to Vedran Leki\'c for providing ellipticity corrections and to Paula Koelemeijer for her question at SEDI 2022 that prompted this research. We thank the editor, Quentin Williams, and reviewers, Andrew Valentine and two anonymous, who have improved this manuscript post-submission. We acknowledge funding from the European Research Council (ERC) under the European Union’s Horizon 2020 research and innovation programme (grant agreement No. 804071-ZoomDeep).

\section*{Open Research}

A table of all modes used, along with their centre frequencies and the errors employed can be found in the supplementary materials and is also available on Zenodo \citep{Database}. To calculate eigenfrequencies and eigenfunctions we used MINEOS version 1.0.2 \citep{Mineos}, published under the GPL2 license.

\DeclareRobustCommand{\VAN}[3]{#3}
\bibliography{bibliography}

\end{document}


\title{Supplementary Materials for:\\ \textbf{Evidence for a kilometre-scale seismically slow layer atop the core-mantle boundary from normal modes}}
\date{}
\author{Stuart Russell\textsuperscript{1,2}, Jessica C.E. Irving\textsuperscript{3}, Lisanne Jagt\textsuperscript{1}, Sanne Cottaar\textsuperscript{1} \\ \footnotesize{\textsuperscript{1} University of Cambridge, United Kingdom}\\ \footnotesize{\textsuperscript{2} Universit\"at M\"unster, Germany}\\ \footnotesize{\textsuperscript{3} University of Bristol, United Kingdom}}

\maketitle

\noindent This document includes:
\begin{enumerate}

    \item A selection of sensitivity kernels for normal modes in different velocity models.

    \item A demonstration of the different behaviour of eigenfrequencies for the solid and molten layers.

    \item An extended results section containing eigenfrequency and measured value comparison for all modes.

    \item A comparison of the results when the background model is changed from EPOC to PREM or STW105.

    \item A comparison of the results for toroidal and spheroidal modes separately.

    \item An exploration of the trade-offs between the strength of the seismic parameters and thickness.

    \item An exploration of the trade-offs with wider D$^{\prime \prime}$ structure and lower mantle radial anisotropy.

    \item An exploration of the effect of changing the CMB radius.

    \item A discussion of the treatment of errors and an exploration of the robustness of the results.
    
    \item Further presentation of results from the grid search.

    \item Further details on the construction of the mode dataset, and a table of the included modes.
    
\end{enumerate}

\pagebreak 

\section{The impact of the model on sensitivity kernels} \label{kernels}

Sensitivity kernels provide a description of the sensitivity of the centre frequency of a particular mode to different parameters at depths within a model. Kernels are calculated from the eigenfunctions outputted from MINEOS \citep{Masters2011} using the formulation from \cite{Dahlen1999}. The eigenfunctions, and therefore the sensitivity kernels, depend on the model that the mode is oscillating within. In this section we demonstrate how the behaviour of selected modes changes between different models.

\begin{figure}[H]
    \centering
    \includegraphics[width=16cm]{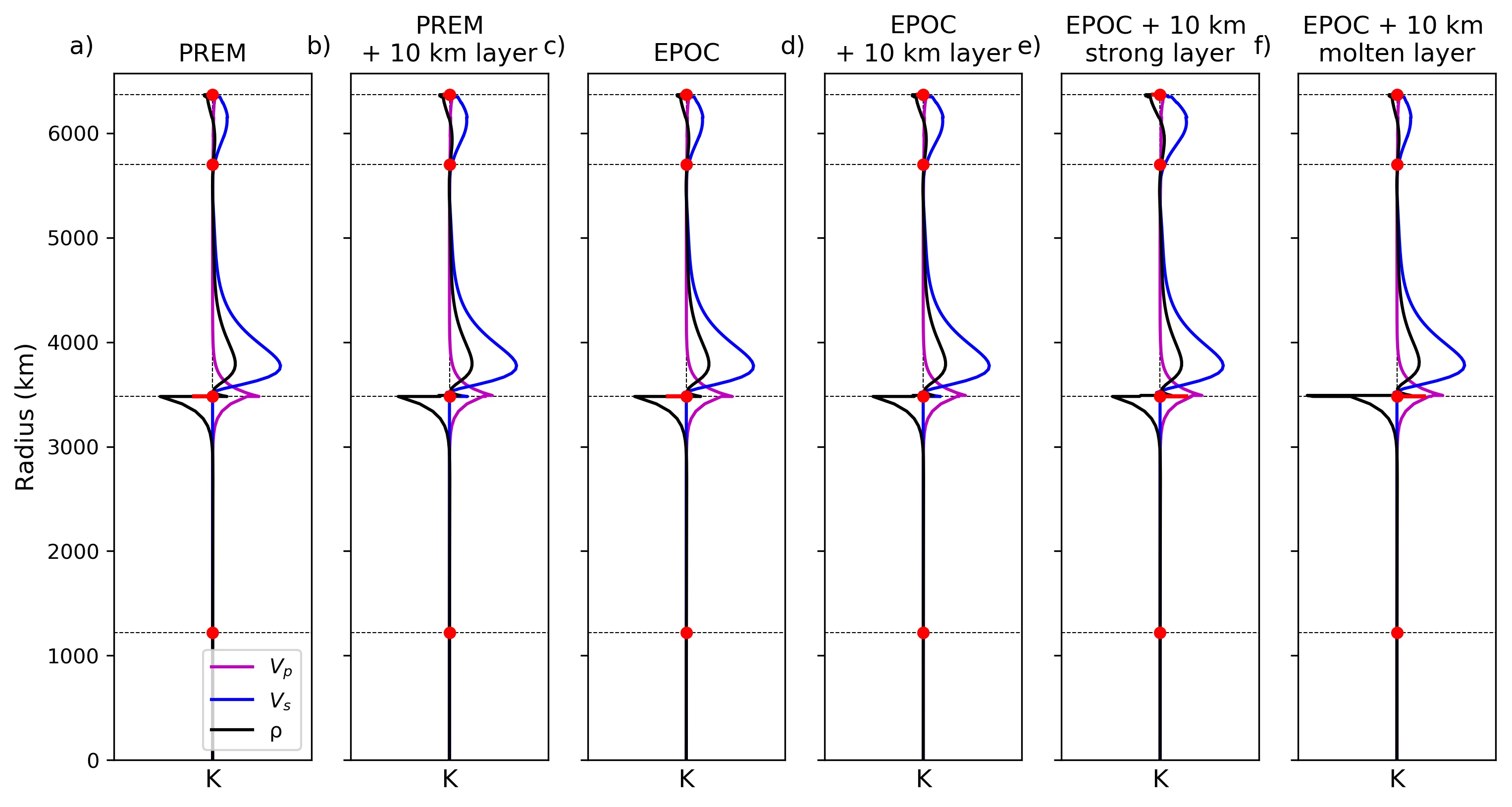}
    \caption{Sensitivity kernels for P-velocity (magenta), S-velocity (blue), density (black) and the displacement of major discontinuities (red) as a function of radius for mode \textsubscript{2}S\textsubscript{16} in (a) PREM, (b) PREM with a 10~km thick solid layer, (c) EPOC, (d) EPOC with a 10~km thick solid layer ($\delta V_{p}$:$-$20\%, $\delta V_{s}$:$-$40\% and $\delta \rho$:$+$50\%) and (e) EPOC with a 10~km thick solid layer where the parameters are twice as extreme, and (f) EPOC with a 10~km thick molten layer ($\delta V_{p}$:$-$33\%, $\delta V_{s}$:$-$100\% and $\delta \rho$:$+$14\%). Horizontal dashed lines represent the major discontinuities: the surface, 660~km, core-mantle boundary (CMB) and inner-core boundary (ICB). Maximum amplitudes are consistent between subplots.}
    \label{kernel_2S16}
\end{figure}

\begin{figure}[H]
    \centering
    \includegraphics[width=16cm]{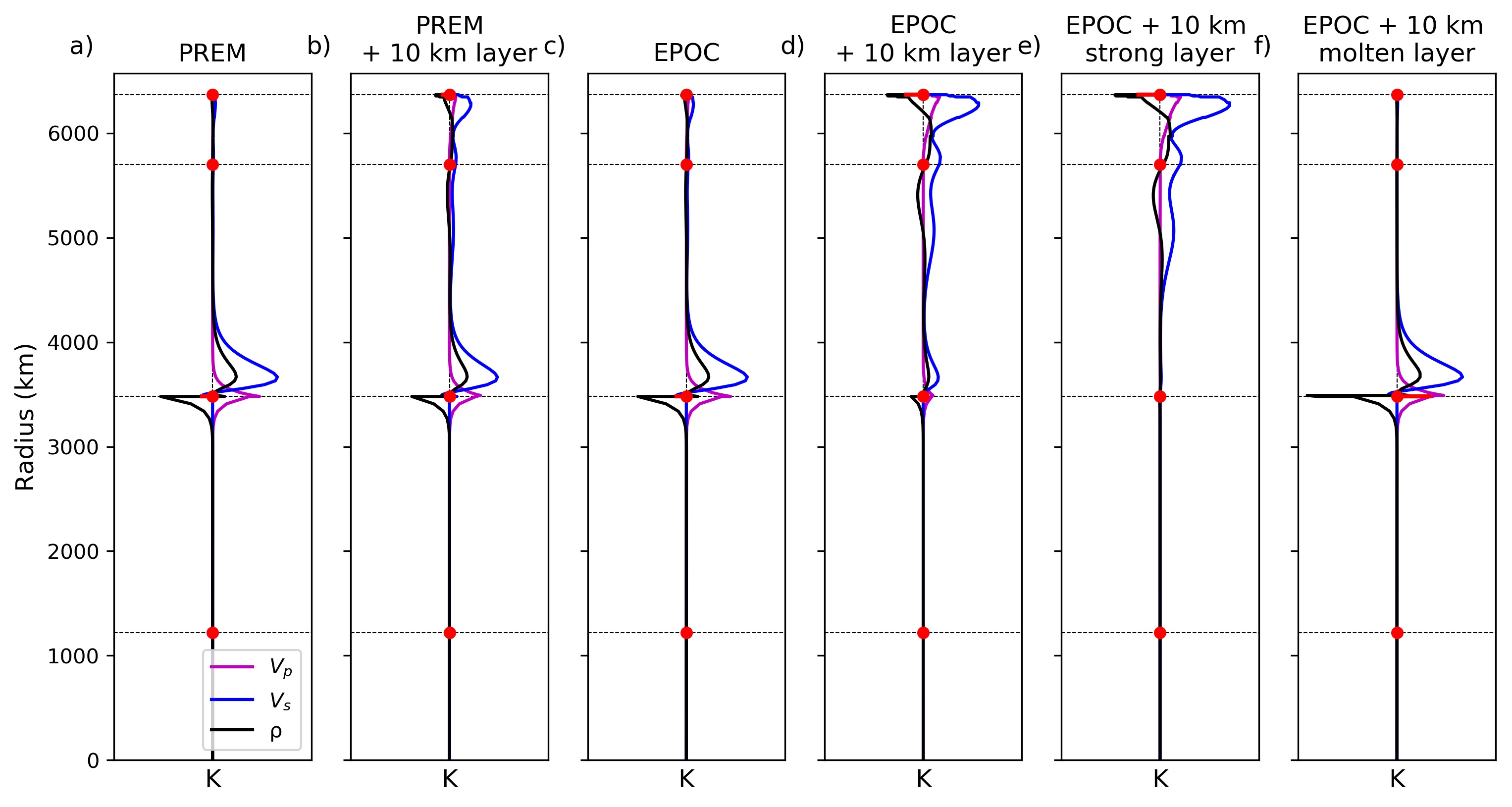}
    \caption{As Figure \ref{kernel_2S16}, for mode \textsubscript{3}S\textsubscript{26}.}
    \label{kernel_3S26}
\end{figure}

\begin{figure}[H]
    \centering
    \includegraphics[width=16cm]{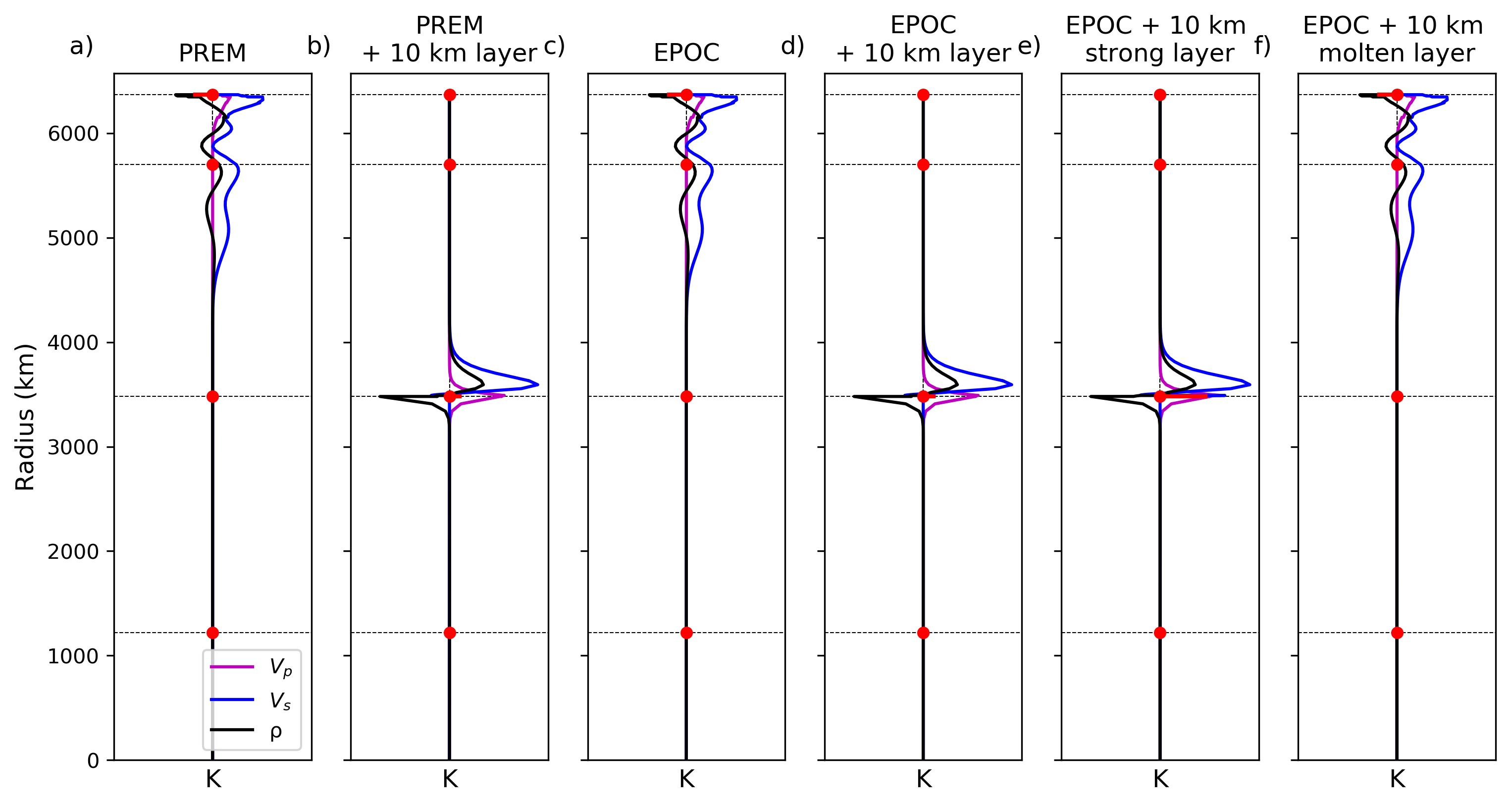}
    \caption{As Figure \ref{kernel_2S16}, for mode \textsubscript{3}S\textsubscript{41}.}
    \label{kernel_3S41}
\end{figure}

\begin{figure}[H]
    \centering
    \includegraphics[width=16cm]{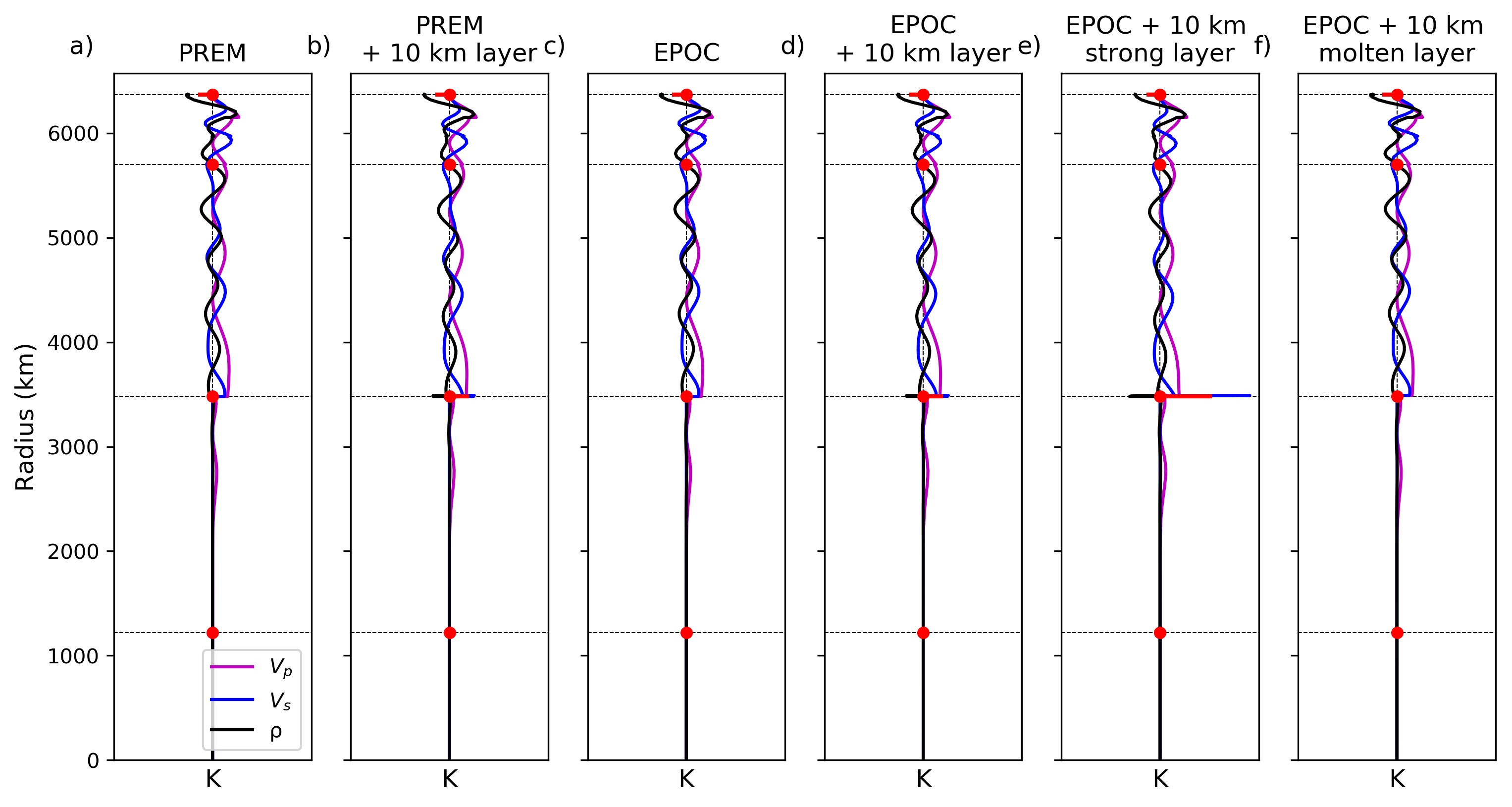}
    \caption{As Figure \ref{kernel_2S16}, for mode \textsubscript{15}S\textsubscript{15}.}
    \label{kernel_15S15}
\end{figure}

\begin{figure}[H]
    \centering
    \includegraphics[width=16cm]{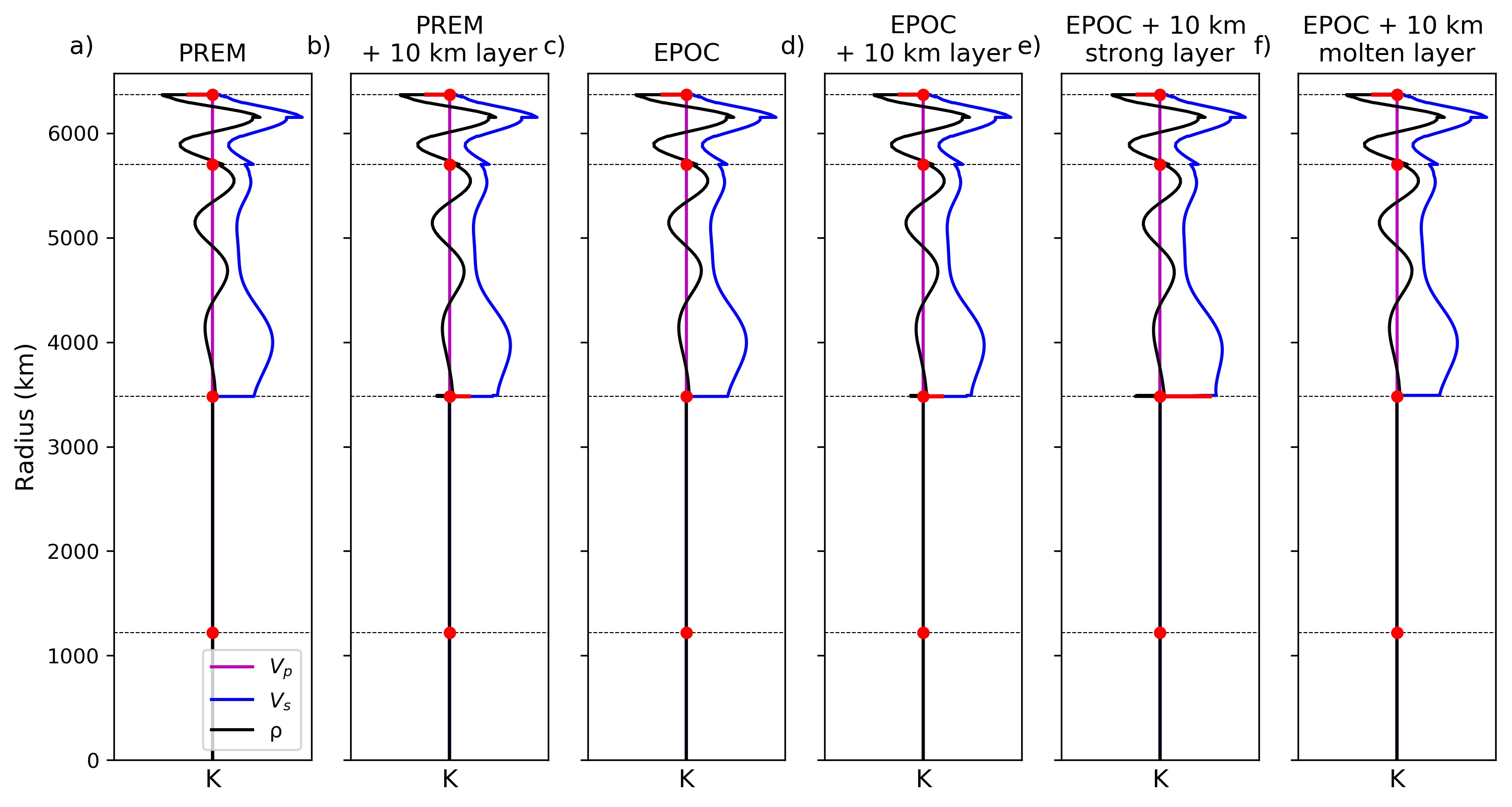}
    \caption{As Figure \ref{kernel_2S16}, for mode \textsubscript{3}T\textsubscript{16}.}
    \label{kernel_3T16}
\end{figure}

Figures \ref{kernel_2S16}-\ref{kernel_3T16} show how the kernels change for five selected modes in both PREM \citep{Dziewonski1981} and EPOC \citep{Irving2018} with and without layers. Some modes, such as \textsubscript{2}S\textsubscript{16} (Figure~\ref{kernel_2S16}) change minimally between different models even when a layer is included. Others drastically change their behaviour, for example \textsubscript{3}S\textsubscript{26} (Figure~\ref{kernel_3S26}) which ceases to be a Stoneley mode \citep{Stoneley1924, Koelemeijer2013} when a layer is included in EPOC, and \textsubscript{3}S\textsubscript{41} (Figure~\ref{kernel_3S41}) which becomes a Stoneley mode in both PREM and EPOC when a layer is included. Even for some modes which do not change significantly throughout most of the Earth, the sensitivity at the CMB is enhanced when a layer is included. This can be clearly seen for \textsubscript{15}S\textsubscript{15} (Figure~\ref{kernel_3S41}) where the S-velocity and density sensitivities are strongly enhanced inside the layer. Layers with more extreme parameters also cause more prominent peaks in the kernels within the layer, as can be seen by comparing panels (d) and (e). \textsubscript{3}T\textsubscript{16} (Figure~\ref{kernel_3T16}) as a toroidal mode has no sensitivity to the outer core, hence does not change between PREM and EPOC, but does have enhanced density sensitivity within the layer. This behaviour of enhanced density sensitivity but not S-velocity is a prevalent feature of the toroidal modes in the dataset.

While the kernels do evolve as layer thickness is increased, drastic changes to the kernels (for example for \textsubscript{3}S\textsubscript{26} and \textsubscript{3}S\textsubscript{41} in Figures \ref{kernel_3S26} and \ref{kernel_3S41}) are the result of `mode swapping', where, due to mode naming conventions, the definition of the mode changes as the model is perturbed greatly. For \textsubscript{3}S\textsubscript{26} and \textsubscript{3}S\textsubscript{41} the modes swap with \textsubscript{2}S\textsubscript{26} and \textsubscript{3}S\textsubscript{41} respectively. Eigenfunctions are assigned to mode names based on assumptions about the Earth and it is not possible to know for certain the `correct' mode to use. In the methodology presented in the main paper, modes for which this is problematic are generally suppressed by having a high model error and do not strongly influence the results. Nevertheless, this demonstrates that the behaviour of some commonly used modes is complex when the model is perturbed very strongly, resulting in difficulties in interpretation.

\pagebreak

\section{Behaviour of individual modes} \label{Behaviour}

\begin{figure}[H]
    \centering
    \includegraphics[width=14cm]{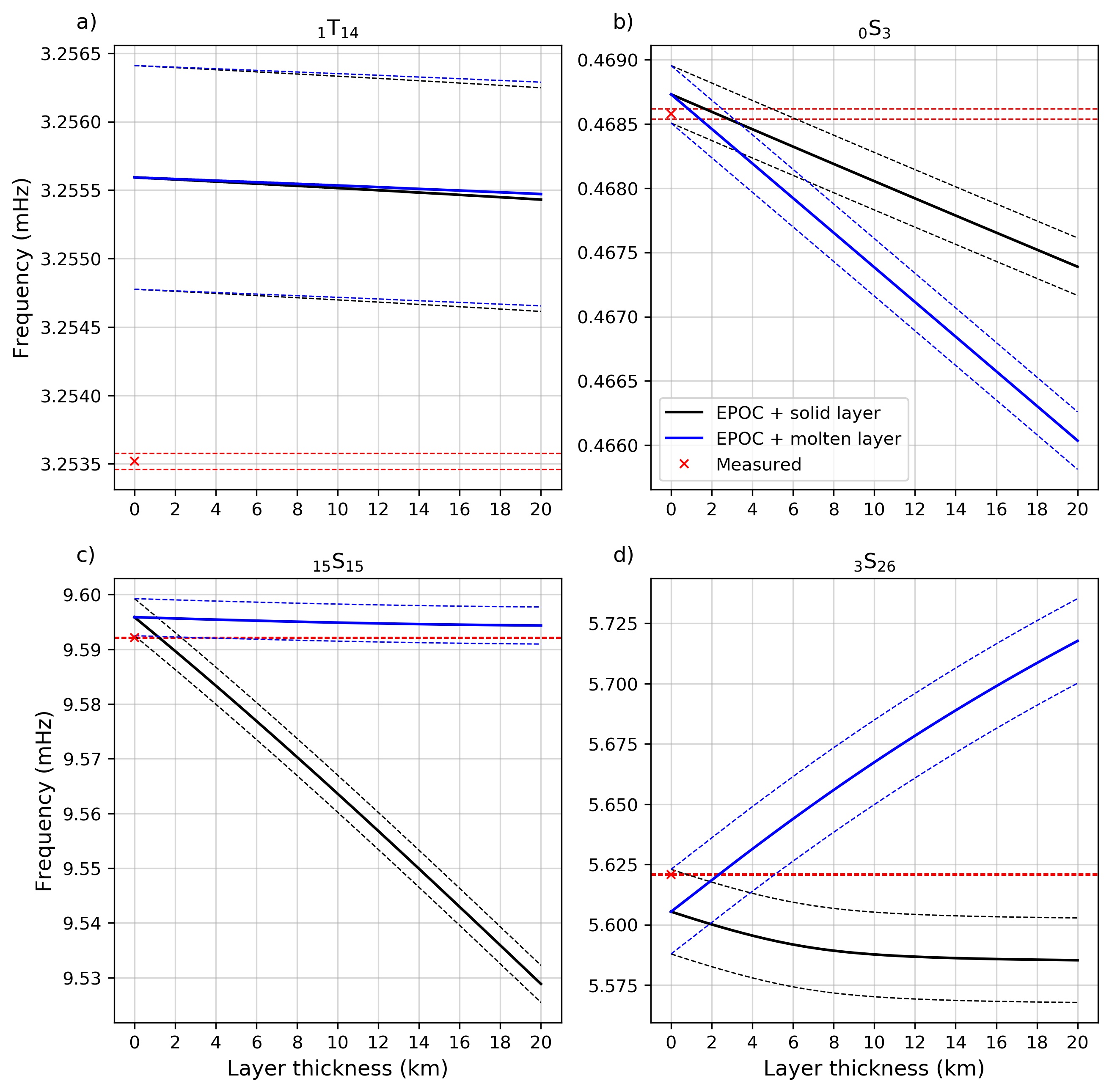}
    \caption{Centre frequency, as a function of layer thickness for four modes: (a) \textsubscript{1}T\textsubscript{14}, (b) \textsubscript{0}S\textsubscript{3}, (c) \textsubscript{15}S\textsubscript{15} and (d)  \textsubscript{3}S\textsubscript{26}. Solid lines are for the solid layer (black, $\delta V_{p}$:$-$20\%, $\delta V_{s}$:$-$40\% and $\delta \rho$:$+$50\%) and molten layer (blue, $\delta V_{p}$:$-$33\%, $\delta V_{s}$:$-$100\% and $\delta \rho$:$+$14\%) with dashed lines showing the model error, $\epsilon_{f}$. The measured centre frequency is plotted as a red cross and the red dashed lines show the measurement error, $\epsilon_{d}$.}
    \label{layer inclusion}
\end{figure}

Figure~\ref{layer inclusion} shows the eigenfrequencies predicted by MINEOS, of four selected modes as a function of increasing layer thickness for both the solid and molten layers. For all modes, the effect of the solid layer is to decrease the eigenfrequency, however the effect of the molten layer varies between modes. The sensitivity of a mode can be approximated by the rate of change of the eigenfrequency with layer thickness; this is the gradient of a given line in Figure~\ref{layer inclusion}. Some modes such as \textsubscript{1}T\textsubscript{14} (Figure~\ref{layer inclusion}a) behave very similarly for both the solid and molten layer, others such as \textsubscript{0}S\textsubscript{3} (Figure~\ref{layer inclusion}b) show similar behaviour but different sensitivities. A number of modes, including \textsubscript{15}S\textsubscript{15} (Figure~\ref{layer inclusion}c), are particularly sensitive only to one layer type. Other modes show vastly different behaviours for the different layers as can be seen for \textsubscript{3}S\textsubscript{26} (Figure~\ref{layer inclusion}d). This behaviour of the solid layer decreasing the eigenfrequency while the molten layer increases it is prevalent amongst the modes of the dataset, with 245 out of 353 modes showing this behaviour.

\pagebreak

\section{Results for all modes} \label{All_modes}

Figures \ref{results0} and \ref{results1} show the change in eigenfrequency for different modes relative to the measured centre frequencies. This is equivalent to Figure 4(a) in the main paper but for all 353 modes used. These modes are also sorted by gradient with the most sensitive first (gradients are given in Table~\ref{dataset_table0} in Section~S8). It can be seen that the width of coloured band is thickest for the most sensitive modes as is expected.

\begin{figure}[H]
    \centering
    \includegraphics[width=16cm]{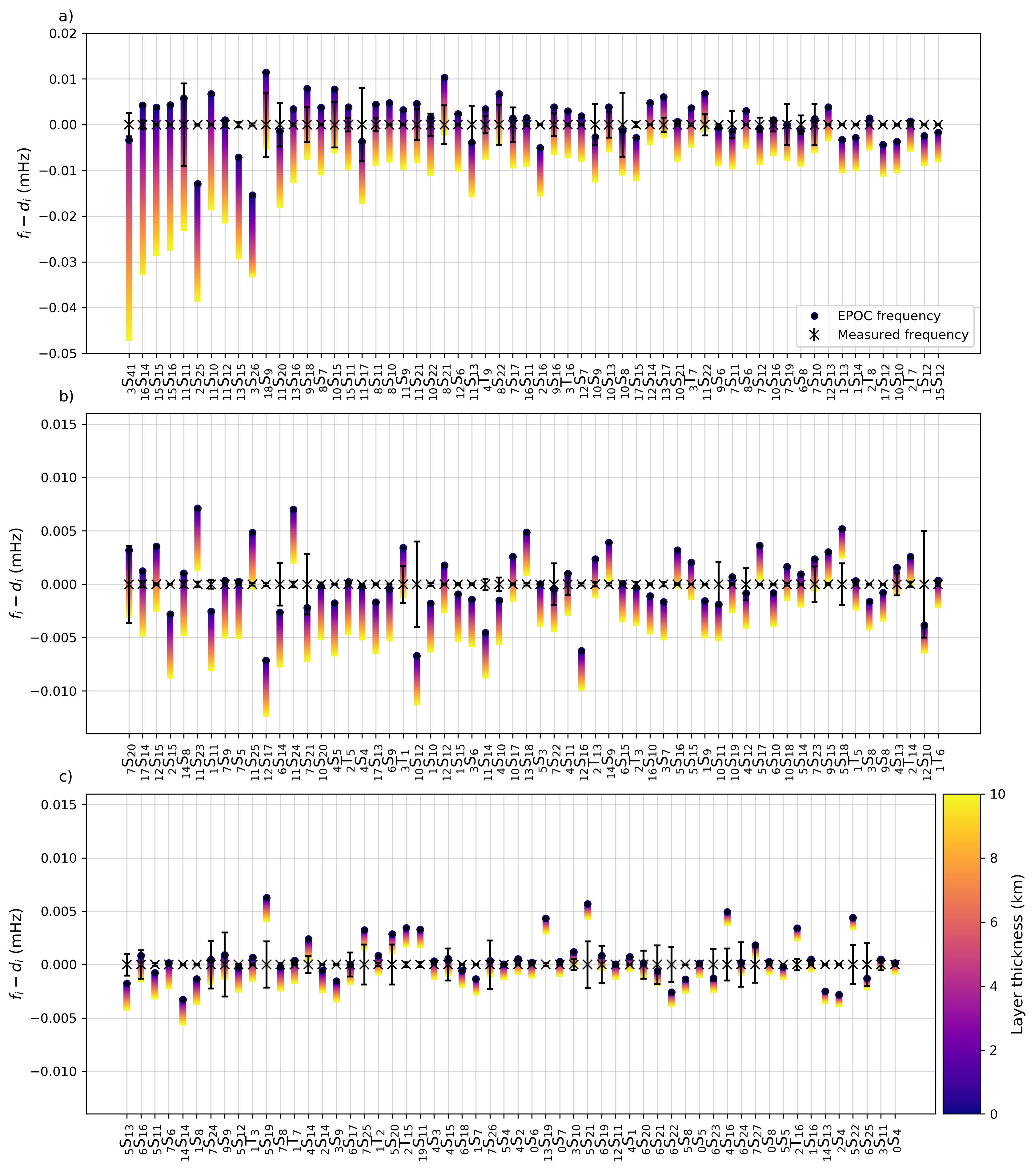}
    \caption{Colours showing the change in eigenfrequencies predicted for different thicknesses of the solid layer in EPOC, $f_{i}$, relative to the measured centre frequencies, $d_{i}$, (black cross) for different layer thicknesses. These plots show the 177 most sensitive modes ordered by gradient with the most sensitive modes to the top left. Note that (a) has a different y-axis scale.}
    \label{results0}
\end{figure}
\setcounter{figure}{6}
\begin{figure}[H]
    \centering
    \includegraphics[width=16cm]{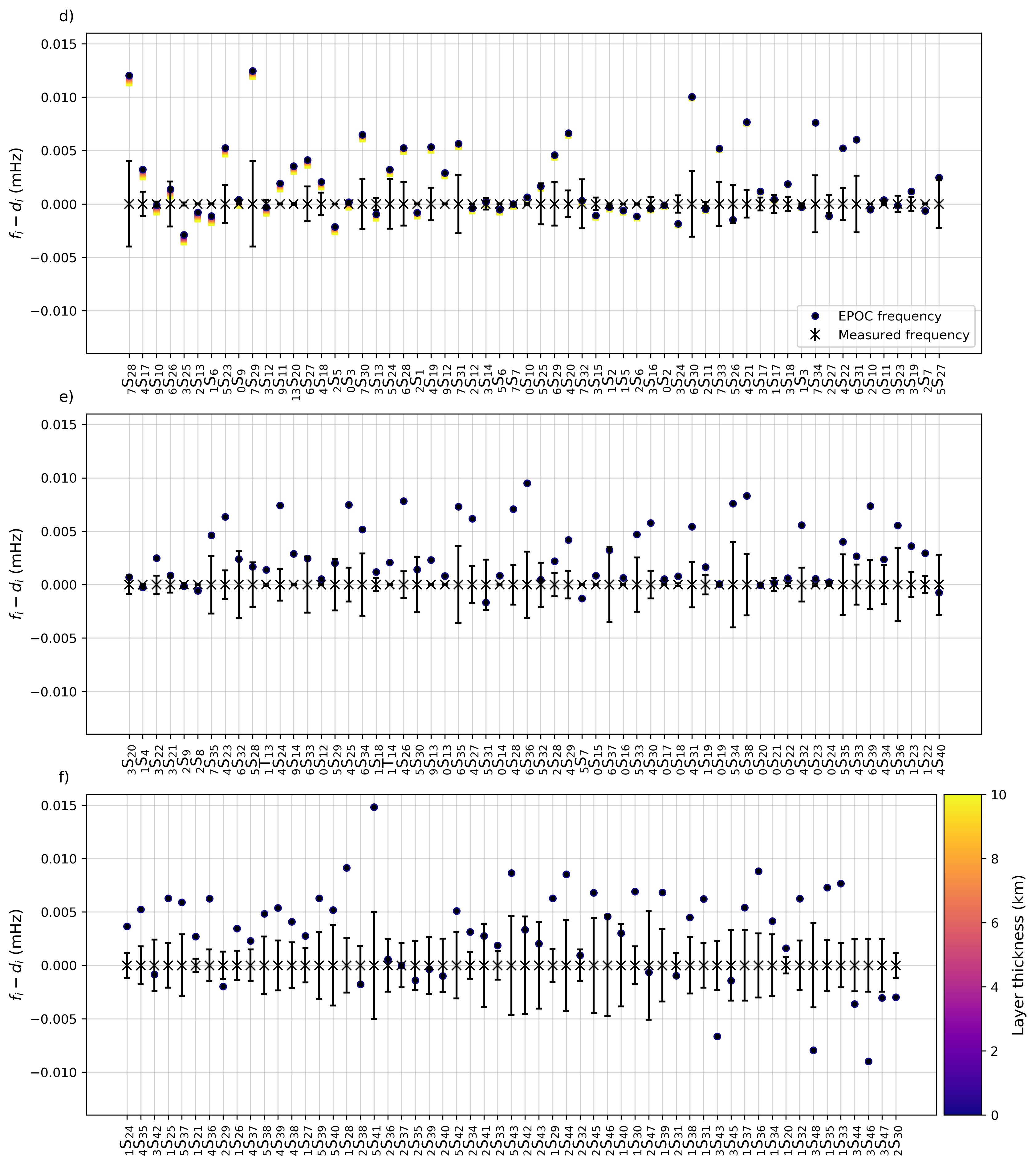}
    \caption{(Continued) Colours showing the change in eigenfrequencies predicted for different thicknesses of the solid layer in EPOC, $f_{i}$, relative to the measured centre frequencies, $d_{i}$, (black cross) for different layer thicknesses. These plots show the 176 least sensitive modes ordered by gradient with the most sensitive modes to the top left. The scales on the y-axes are consistent with panels (b) and (c).}
    \label{results1}
\end{figure}

\pagebreak
\section{The effect of the choice of background model} \label{background_model}

\begin{figure}[H]
    \centering
    \includegraphics[width=16cm]{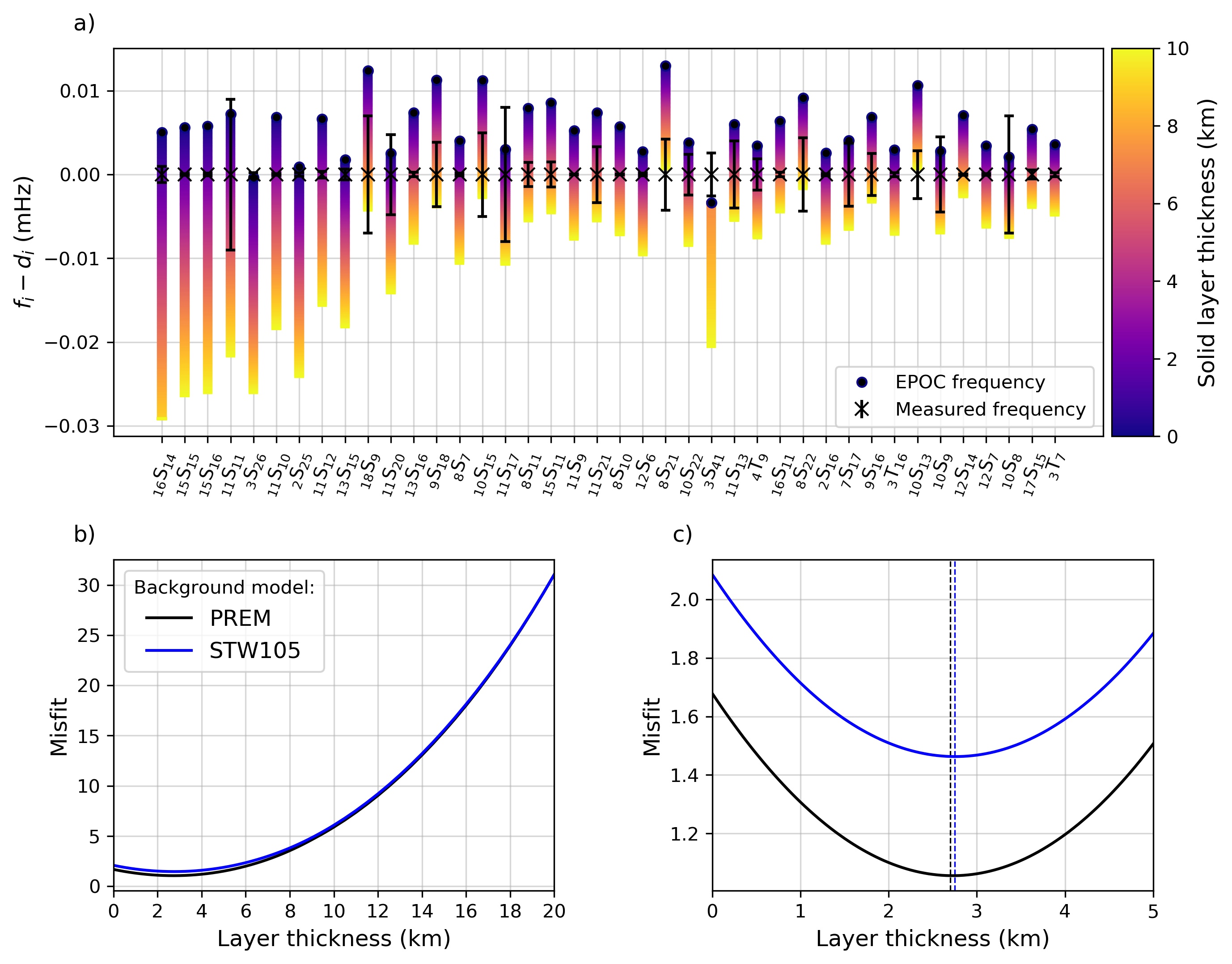}
    \caption{As Figure 2 in the main paper but testing the effect of the background model. (a) Colours showing the change in eigenfrequencies predicted for different thicknesses of the solid layer in PREM, $f_{i}$, relative to the measured centre frequencies, $d_{i}$, (black cross) for different layer thicknesses. Measurement errors, $\epsilon_{d}$, are the black error bars. These are the 40 most sensitive modes ordered by their sensitivity to the inclusion of a layer. (b) and (c) show the misfit, $M$, as a function of layer thickness between 0~-~20~km and 0~-~5~km,  respectively, for solid layers in PREM (black) and STW105 (blue). The vertical dotted line marks the minima in $M$.}
    \label{prem_results}
\end{figure}

Figure~\ref{prem_results} shows results in the same format as Figure 4 of the main paper but for when PREM is used as the background model as opposed to EPOC. We also test STW105 \citep{Kustowski2008} as a background model. PREM and EPOC differ in the outer core, while PREM and STW105 differ in the upper mantle. It should be noted that the most sensitive $n$ modes are not the same for PREM, STW105 and EPOC due to differences in mode sensitivity for the different models, hence why the modes appearing in panel (a) differ between models.

For PREM and STW105 the best-fitting layer thickness is 2.7~km and 2.75~km, respectively, which substantially thicker than the preferred thickness of 1.9~km for EPOC. That there is a difference between these two models and EPOC indicates that there is trade-off between structures above and below the CMB. The differences in the outer core in EPOC, notably a lower velocity immediately below the CMB, can account for some of the misfit between PREM, STW105, and the observed centre frequencies. That the PREM and STW105 solutions are very similar indicates that there is little trade-off with upper mantle structure.

\section{A comparison of toroidal and spheroidal mode results} \label{toroidal}

Toroidal and spheroidal modes are very different in their behaviour and sensitivities - spheroidal modes have radial and tangential motion analogous to P-SV motion, while toroidal modes are purely tangential and are analogous to SH motion. Toroidal modes therefore have no sensitivity to P-velocity and also have no displacement or sensitivity in fluid parts of the Earth. When studying the CMB this is notable as toroidal mode energy must be zero below the CMB.

Toroidal mode measurements are difficult to make and in the last two decades only one dataset of toroidal mode measurements has been published \citep{Schneider2021}; our dataset contains the 19 modes from \cite{Schneider2021}. Toroidal modes are sensitive to the inclusion of a solid CMB layer and there are three toroidal mode in the 50 most sensitive modes, and a total of nine in the 100 most sensitive modes.

\begin{figure}[H]
    \centering
    \includegraphics[width=16cm]{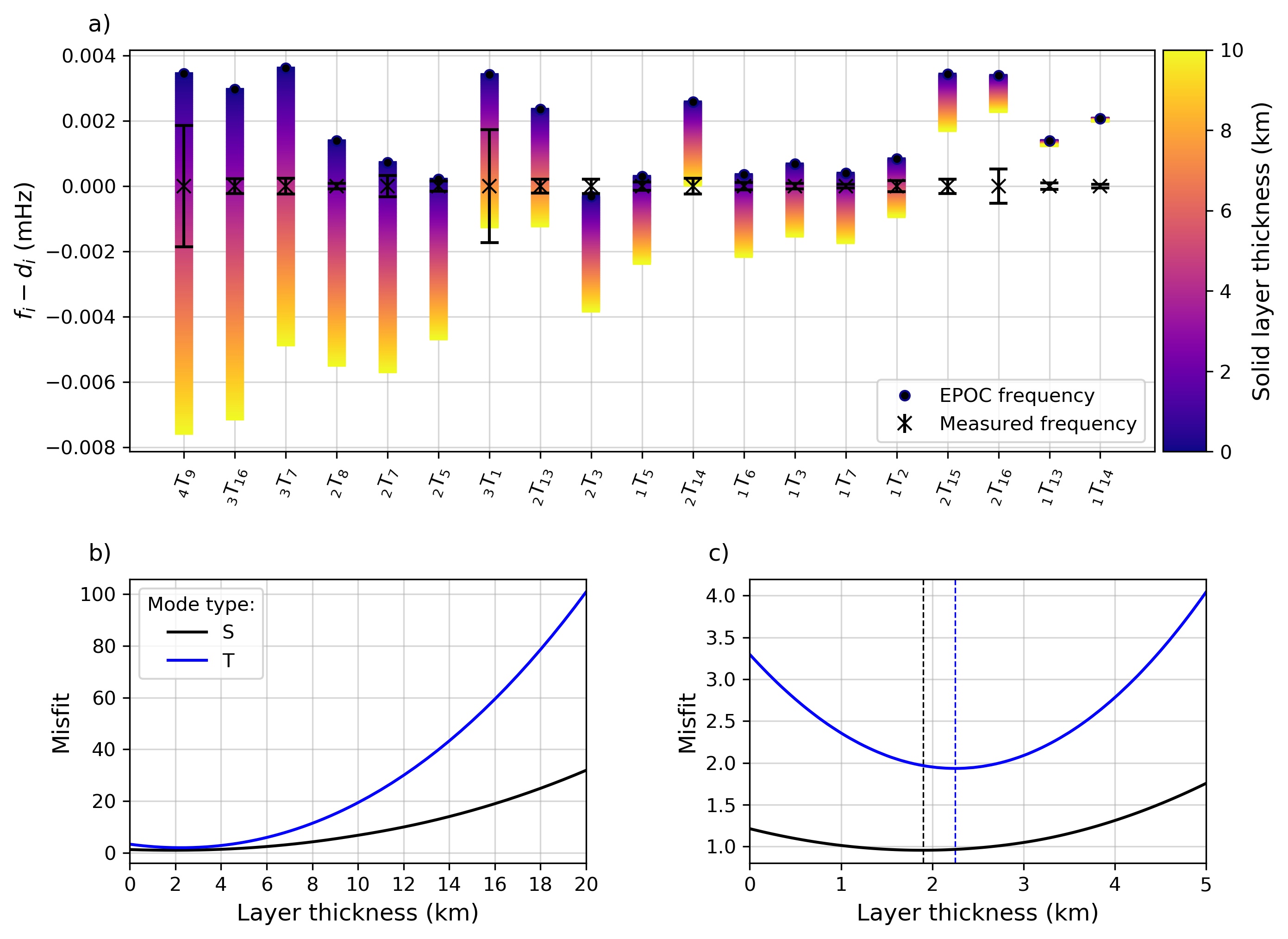}
    \caption{As Figure 2 in the main paper but highlighting the behaviour of toroidal modes. (a) Colours showing the change in eigenfrequencies predicted for different thicknesses of the solid layer in EPOC, $f_{i}$, relative to the measured centre frequencies, $d_{i}$, (black cross) for different layer thicknesses. Measurement errors, $\epsilon_{d}$, are the black error bars. These are the 40 most sensitive modes ordered by their sensitivity to the inclusion of a layer. (b) and (c) show the misfit, $M$, as a function of solid layer thickness between 0~-~20~km and 0~-~5~km, respectively, for spheroidal (black) and toroidal (blue). The vertical dotted line marks the minima in $M$.}
    \label{toroidal_results}
\end{figure}

Figure~\ref{toroidal_results}a shows the results for individual toroidal modes, many of which favour non-zero layer thickness. Figures \ref{toroidal_results}b-c show the results for toroidal and spheroidal modes separately. For toroidal modes there is no difference in the results between PREM and EPOC owing to the fact that toroidal modes have no outer core sensitivity. The toroidal modes have a preferred layer thickness of 2.35~km, thicker than the spheroidal preference of 1.8~km.  The reason for this difference is unclear but may be related to unmodelled lower mantle anisotropy as the PREM (and therefore EPOC) lower mantle is isotropic, as is the layer we include. Alternatively this discrepancy could result from sensitivity to different parameters - toroidal modes are more sensitive to density at the CMB and have no sensitivity to P-velocity.

\pagebreak

\pagebreak
\section{Trade-off with thickness and layer strength} \label{tradeoff}

\begin{figure}[H]
    \centering
    \includegraphics[width=16cm]{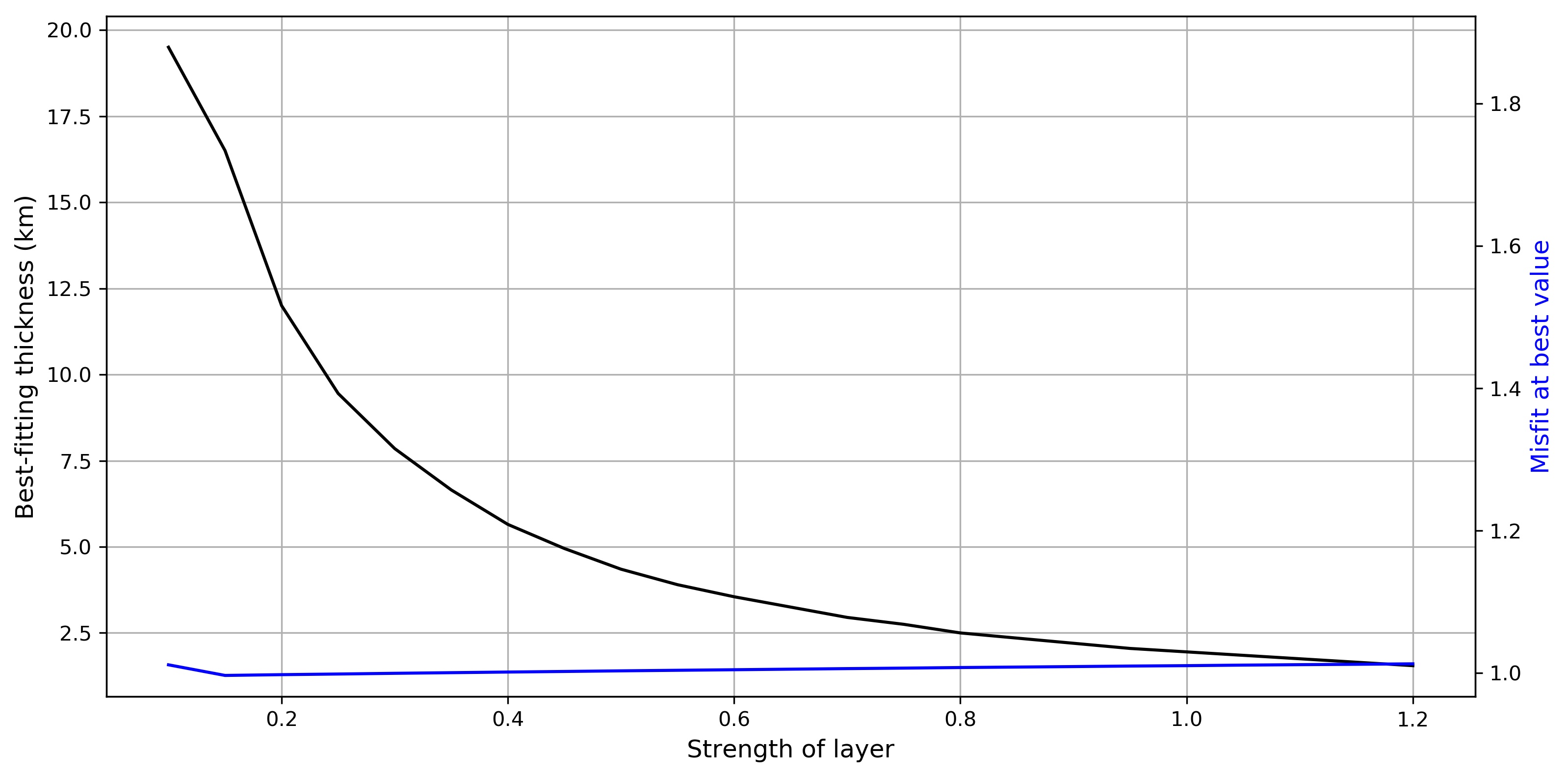}
    \caption{Best-fitting layer thickness (black) and the misfit at this value (blue) as a function of the strength of the layer. Strength is defined such that a strength of 0 is PREM parameters while a strength of 1 are the solid layer parameters given in the main paper. Therefore strength values between 0 and 1 are linearly intermediate between PREM and the layer, while strengths greater than 1 have parameters more extreme than the original layer. The y-axis for the misfit is scaled the same as Figure 4c in the main paper.}
    \label{deviation}
\end{figure}

Layer parameters were also varied in severity from those of the solid layer in the main paper. Figure~\ref{deviation} shows the trade-off between the strength of the seismic parameters of a layer and its best-fitting thickness, as well as the misfit at this best-fitting value for the solid layer. Less extreme layers are better fitting when they are thicker; this is expected but represents a significant source of non-uniqueness in the solution, especially as the misfit does not change significantly across the range of strengths tested. Thicker and less extreme layers are marginally better fitting but this is insignificant compared to the effect of including a velocity reduction at all.

\pagebreak

\section{Trade-offs with D$^{\prime \prime}$ structure and lower mantle anisotropy} \label{LM_Dpp}

\begin{figure}[H]
    \centering
    \includegraphics[width=16cm]{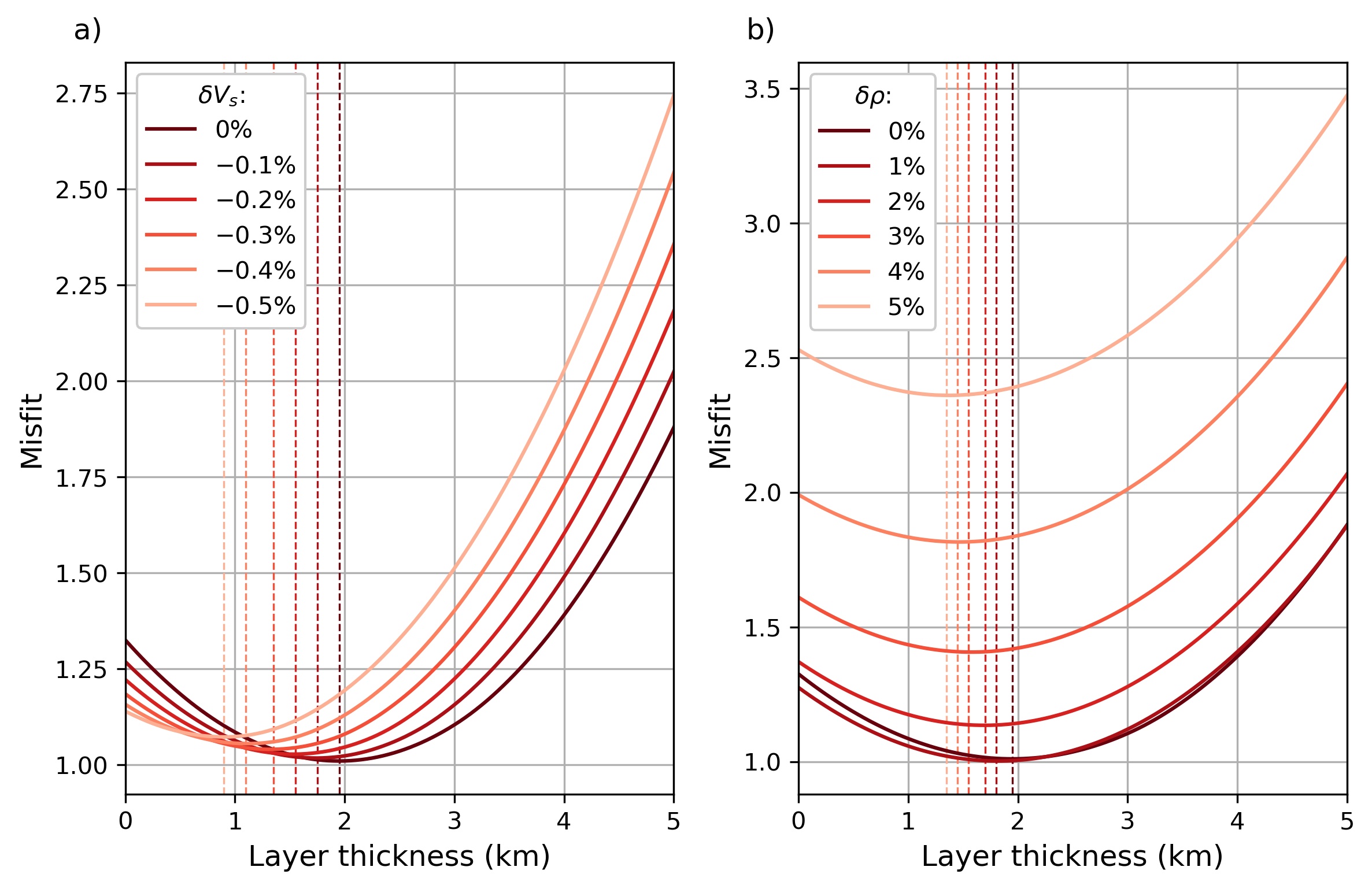}
    \caption{Misfit curves as a function of layer thickness as the background structure of D$^{\prime \prime}$ is varied uniformly for (a) S-velocity and (b) density. Dashed vertical lines mark the best fitting layer thicknesses.}
    \label{Dpp tradeoff}
\end{figure}

We also test the effect of varying structure in the wider lowermost mantle to examine how this trades-off with the preferred layer thickness. First we perform a simple test with wider D$^{\prime \prime}$ structure by decreasing the S-velocity of D$^{\prime \prime}$ uniformly from 2741~km to 2891~km in EPOC and recalculate how the misfit varies with layer thickness for this background model. We also perform the same test for increased density. We include these tests simply to highlight the trade-offs in our results, motivated by studies that have suggested that D$^{\prime \prime}$ may be dense or have reduced S-velocity relative to PREM \citep[e.g.][]{Montagner1996, deWit2015}. The results are shown in Figure~\ref{Dpp tradeoff}.

It is apparent that trade-offs with wider D$^{\prime \prime}$ structure do exist, however in all cases tested the best-fitting results were still a layer of the same order of magnitude thickness. The results for the $V_{s}$ reduction are especially interesting as without considering a thin layer, a reduced $V_{s}$ throughout D$^{\prime \prime}$ is better-fitting than PREM, yet when a thin layer is allowed the best-fitting solution is to concentrate the velocity reduction into this layer. These tests are not comprehensive and the results should not be over-interpreted, however they do highlight the trade-offs that exist between thin and D$^{\prime \prime}$-wide properties in the lowermost-mantle.

\pagebreak

We also perform simplistic tests with lower mantle radial anisotropy. It is well known that parts of the upper mantle are radially anisotropic \citep[e.g. see][]{Dziewonski1981}, and there are now several studies that suggest lower mantle radial anisotropy based on normal mode observations \citep[e.g.][]{Montagner1996, deWit2015}. We perform two tests, one with a radially anisotropic D$^{\prime \prime}$ and one with radial anisotropy present throughout the lower mantle. These tests are again overly simplistic and should not be over-interpreted as they are only performed to assess whether anisotropy has a significant affect on our results.

We first test the effect of an anisotropic D$^{\prime \prime}$ with anisotropy in the $\phi$ and $\eta$ parameters \citep[as per the ak135f line from Figure 12 of][]{Montagner1996}. A more recent study has found similar anisotropic parameters for D$^{\prime \prime}$, albeit with great posterior uncertainty \citep{deWit2015}. The aim of this test is not to assess the viability of these models, but simply to assess whether possible D$^{\prime \prime}$ impacts our results. We therefore include uniform anisotropy of $\phi=1.04$ and $\eta=1.02$ throughout D$^{\prime \prime}$, keeping the background model as EPOC. We calculate the best-fitting layer thickness, which for this model is 1.85~km, 100~m less than for isotropic EPOC.

We also test radial anisotropy throughout the lower mantle in the $\phi$ and $\eta$ parameters \citep{deWit2015}. Following Figure 1 of \cite{deWit2015}, we weakly reduce $\phi$ through the lower mantle and we reduce $\eta$ weakly in the depth range 1450~-~1900~km and more strongly below 1900~km. In order to test lower mantle anisotropy independently to D$^{\prime \prime}$, and given the large posterior uncertainties in D$^{\prime \prime}$ anisotropy, we keep D$^{\prime \prime}$ isotropic. We again calculate the best-fitting layer thickness, which is 1.8~km, 150~m less than for isotropic EPOC.

Both of the tests described above indicate that anisotropy in the lower mantle does affect our results, but not significantly enough to undermine the conclusions presented in the main paper. 

\pagebreak

\section{The effect of altering the radius of the CMB} \label{CMB_radius}

\begin{figure}[H]
    \centering
    \includegraphics[width=16cm]{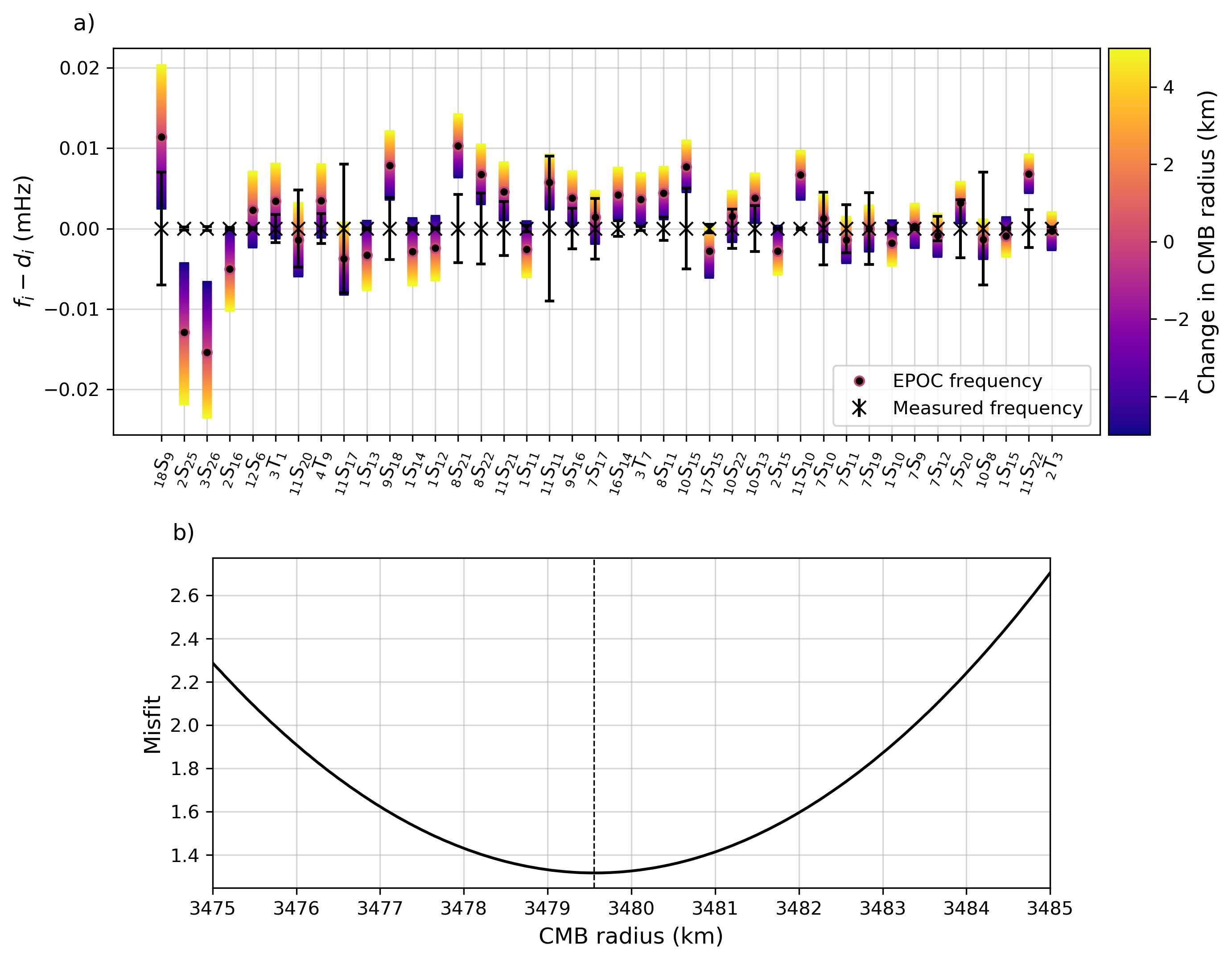}
    \caption{Similar to Figure 2 in the main paper but for models with testing the effect of changing CMB radii. (a) Colours showing the change in eigenfrequencies predicted for different CMB radii in EPOC, $f_{i}$, relative to the measured centre frequencies, $d_{i}$, (black cross) for different layer thicknesses. Measurement errors, $\epsilon_{d}$, are the black error bars. These are the 40 most sensitive modes ordered by their sensitivity to moving the CMB. (b) misfit, $M$, plotted as a function of CMB radius. The vertical dotted line marks the minimum in $M$.}
    \label{CMB_results}
\end{figure}

Given that the misfit is lowest with a velocity reduction above the CMB, we also test whether the same result can be achieved by altering the CMB radius. In PREM and EPOC the CMB radius is 3480~km and we vary it between 3475~km and 3485~km by compressing or expanding the outer core or lower mantle up to 660~km. The results are shown in Figure~\ref{CMB_results}. The best-fitting CMB radius is 3479.55~km, closer to the value of ak135 \citep{Kennett1995} than of PREM, however the misfit is not as low as introducing a low velocity layer indicating that altering the CMB radius alone cannot explain our results. It is possible that there are trade-offs between CMB radius, upper outer core structure, and any slow layering atop the CMB, however a full exploration of this is beyond the scope of this study.

\pagebreak

\section{Treatment of Errors} \label{errors}

\cite{Akbarashrafi2018} have shown that published uncertainties of splitting function measurements, and by extension of centre frequencies, are likely significantly underestimated. This is especially true for uncertainties estimated by cross-validation which do not account for errors originating outside of the measurement process, such as theoretical approximations, which may introduce systematic errors.

In the dataset used for this study, uncertainties from publications that estimate their uncertainties by cross-validation \citep{Deuss2013,Koelemeijer2013,Schneider2021} are especially small, and for the purpose of this section will be referred to as DKS modes. Within the dataset, the mean measurement error on DKS modes is a factor of 15.875 times smaller than from other sources. This is primarily because these uncertainties are defined differently, however these measurements are also more recent and are made with more, and better quality, data. Furthermore in the dataset, the DKS modes are primarily lower $l$ (all modes where $l$ \textless\ 9 are DKS modes), with smaller magnitude centre frequencies which are typically better constrained (the mean centre frequency of DKS modes is 4.56~mHz in contrast to 6.80~mHz for other modes). It is therefore not true that the uncertainties are underestimated by a factor of 15.875, but rather by an unknown smaller value. We chose not to increase the measurement errors by an arbitrary amount and instead use the published values regardless. We investigate the validity of this below.

\begin{figure}[H]
    \centering
    \includegraphics[width=16cm]{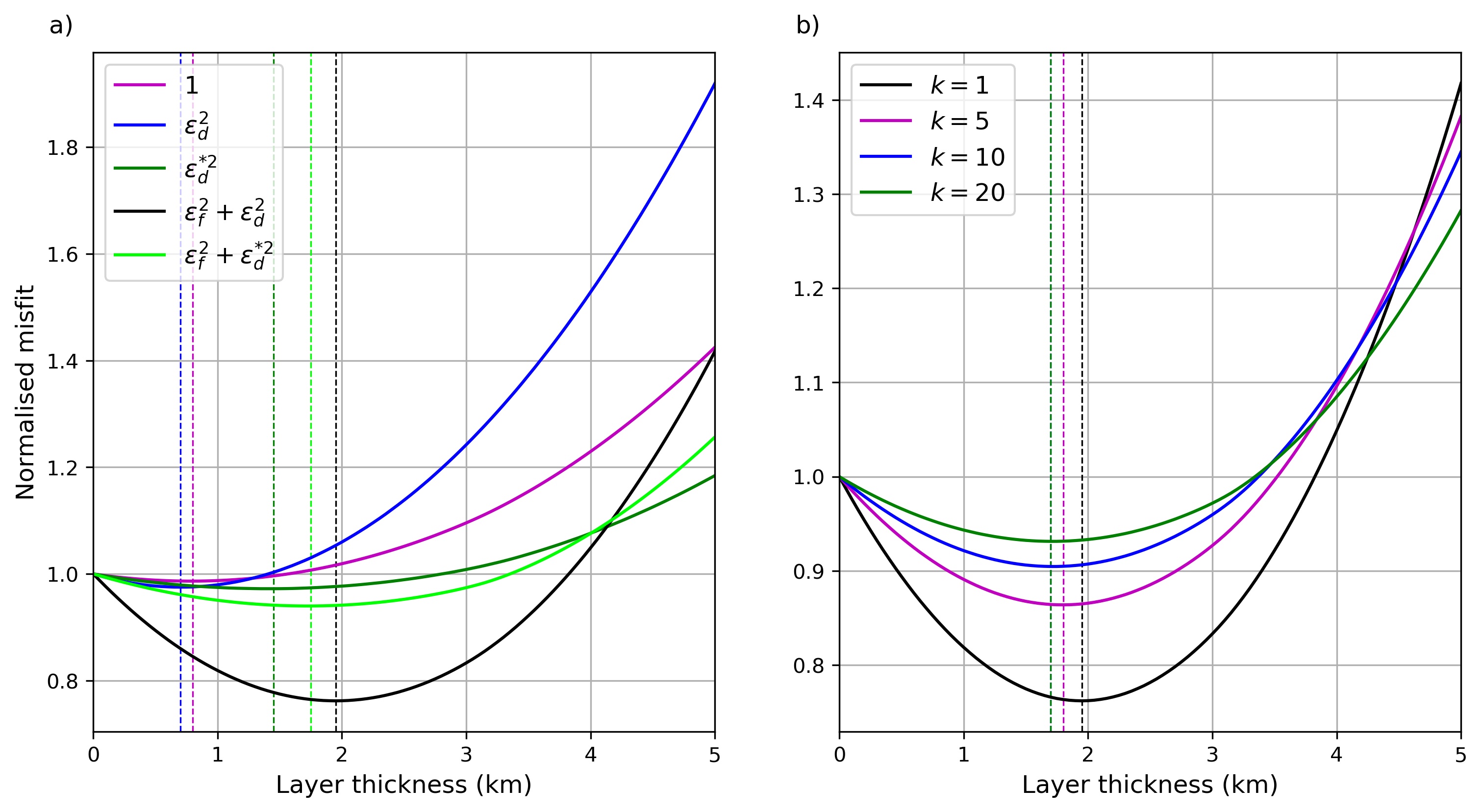}
    \caption{The misfit as a function of layer thickness when different error combinations are used. Each line uses a different denominator for the misfit equation as given in the legend and is normalised to the EPOC misfit. $\epsilon_{d}^{*}$ indicates that the measurement errors on DKS modes only are increased by a factor of 15.875. In (b), the denominator used in the misfit equation is $\epsilon_{f}^{2} + (k\epsilon_{d})^{2}$.}
    \label{all_errors}
\end{figure}

Firstly we test how the best-fitting layer thickness for the solid layer changes when different combinations of errors are used. We do this by changing the denominator of the misfit equation in the main paper. The results are shown in Figure \ref{all_errors}. We test the equation with no weighting by errors (denominator equal to 1), weighting by only the measurement error and weighting by a combination of both measurement and model error. For the final two scenarios, we also test with the measurement errors on DKS modes multiplied by 15.875, which is then an over-estimation of the uncertainty. In all cases, the best-fitting thickness is non-zero and the underestimated uncertainties do not change the results significantly enough to change the conclusions or interpretations.

\pagebreak

\begin{figure}[H]
    \centering
    \includegraphics[width=16cm]{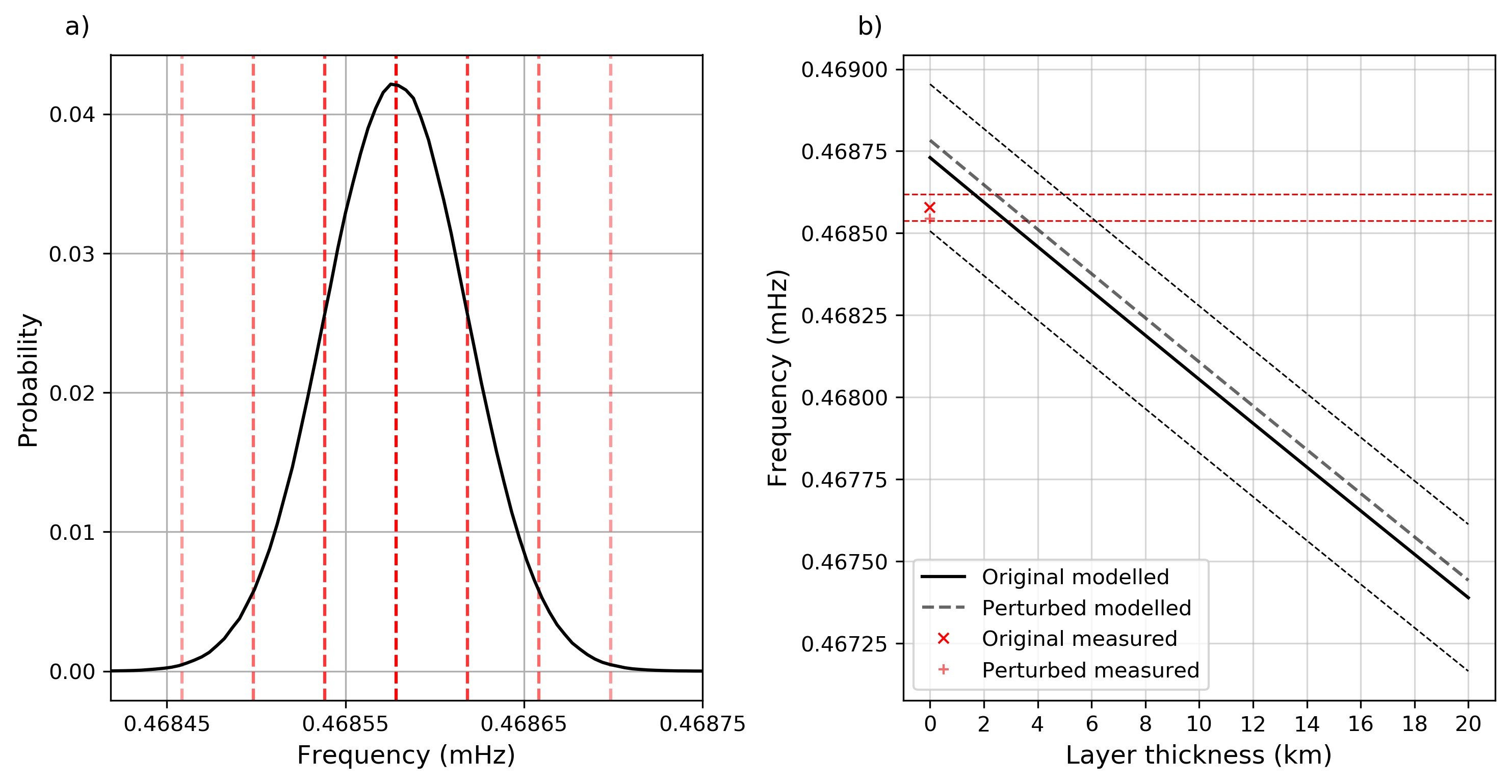}
    \caption{Demonstration of how the measured and modelled centre frequencies are perturbed for mode \textsubscript{0}S\textsubscript{3}. (a) a Gaussian distribution centred on the centre frequency of \textsubscript{0}S\textsubscript{3} with standard deviation equal to the measurement error. Vertical red dashed lines mark 0, 1, 2, and 3 standard deviations. (b) an example demonstrating how the measured and modelled centre frequencies are perturbed.}
    \label{err_pert_demo}
\end{figure}

Even if the underestimated uncertainties do not change the results, they will affect the robustness of them. In order to see how these propagate through our method we perturb both the measured and modelled centre frequencies randomly based on a Gaussian with standard deviation equal to the measurement and model errors, respectively. A demonstration of this is given in Figure \ref{err_pert_demo}. We also repeat this where the measurement errors on DKS modes are multiplied by 15.875, and where all measurement errors are increased uniformly. For each scenario we perform 1,000,000 iterations and assess the spread in the resulting best-fitting thicknesses of the solid layer.

\begin{figure}[H]
    \centering
    \includegraphics[width=12cm]{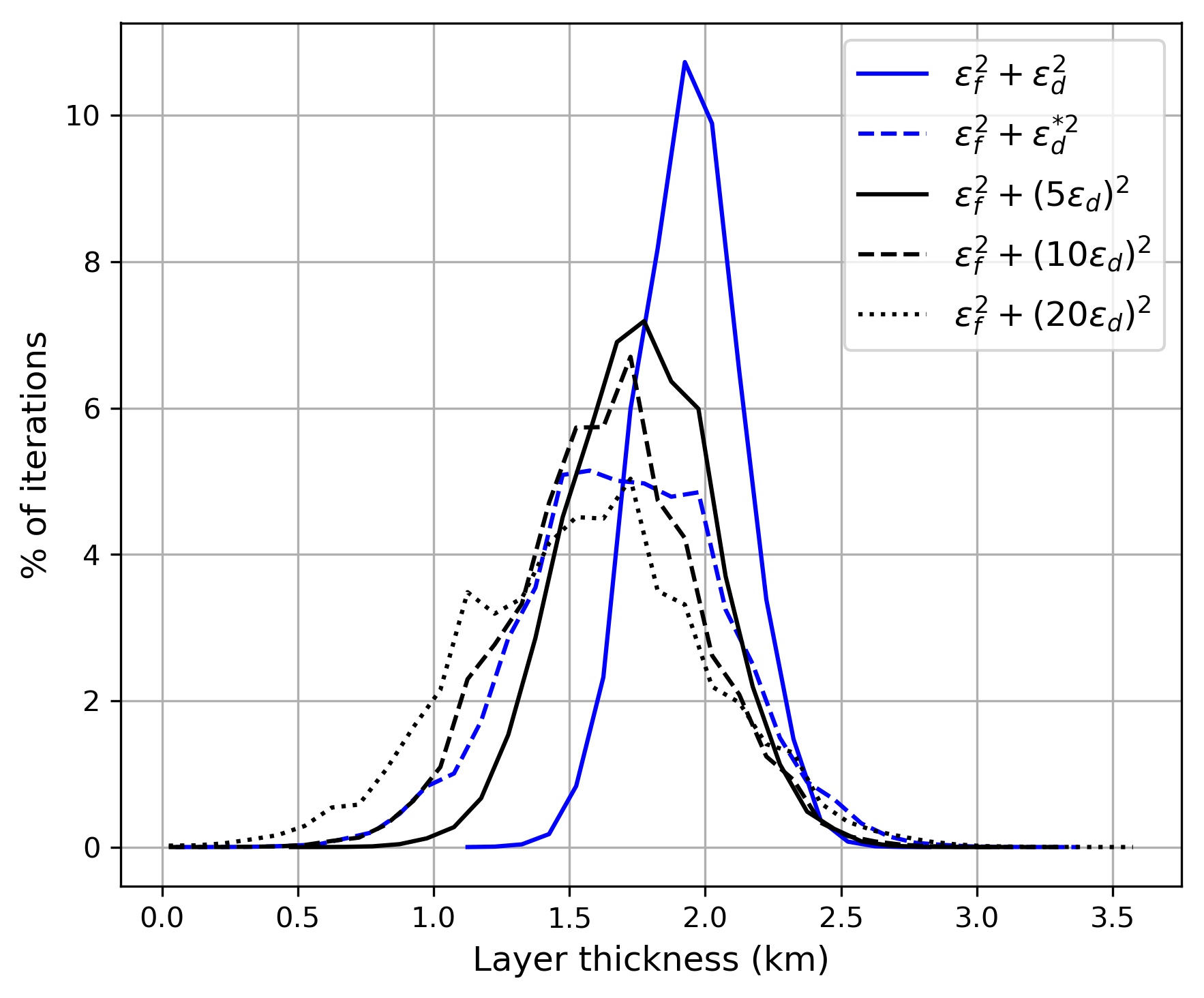}
    \caption{The results of the error propagation analysis. Curves show the distribution of the best-fitting layer thicknesses after one million iterations. Each curve is for a different denominator of the misfit function, given in the legend. $\epsilon_{d}^{*}$ indicates that the measurement errors on DKS modes only are increased by a factor of 15.875.}
    \label{err_prop}
\end{figure}

The results of this analysis are shown in Figure \ref{err_prop}. With the original measurement errors, the best-fitting thickness varies considerably, but is always thicker than 1~km. When the measurement errors on DKS modes are increased, the best-fitting thickness varies more, as would be expected, but remains non-zero in all but 3 out of one million iterations. When all measurement errors are increased by a factor of 10, only one in one million iterations preferred no layer. When they are increased by a factor of 20, 0.03\% of iterations preferred no layer. We must therefore conclude that the modes robustly favour the presence of a layer regardless of any underestimation of the uncertainties. 

\pagebreak

\section{Grid search results} \label{parameters}

\begin{figure}[H]
    \centering
    \includegraphics[width=16cm]{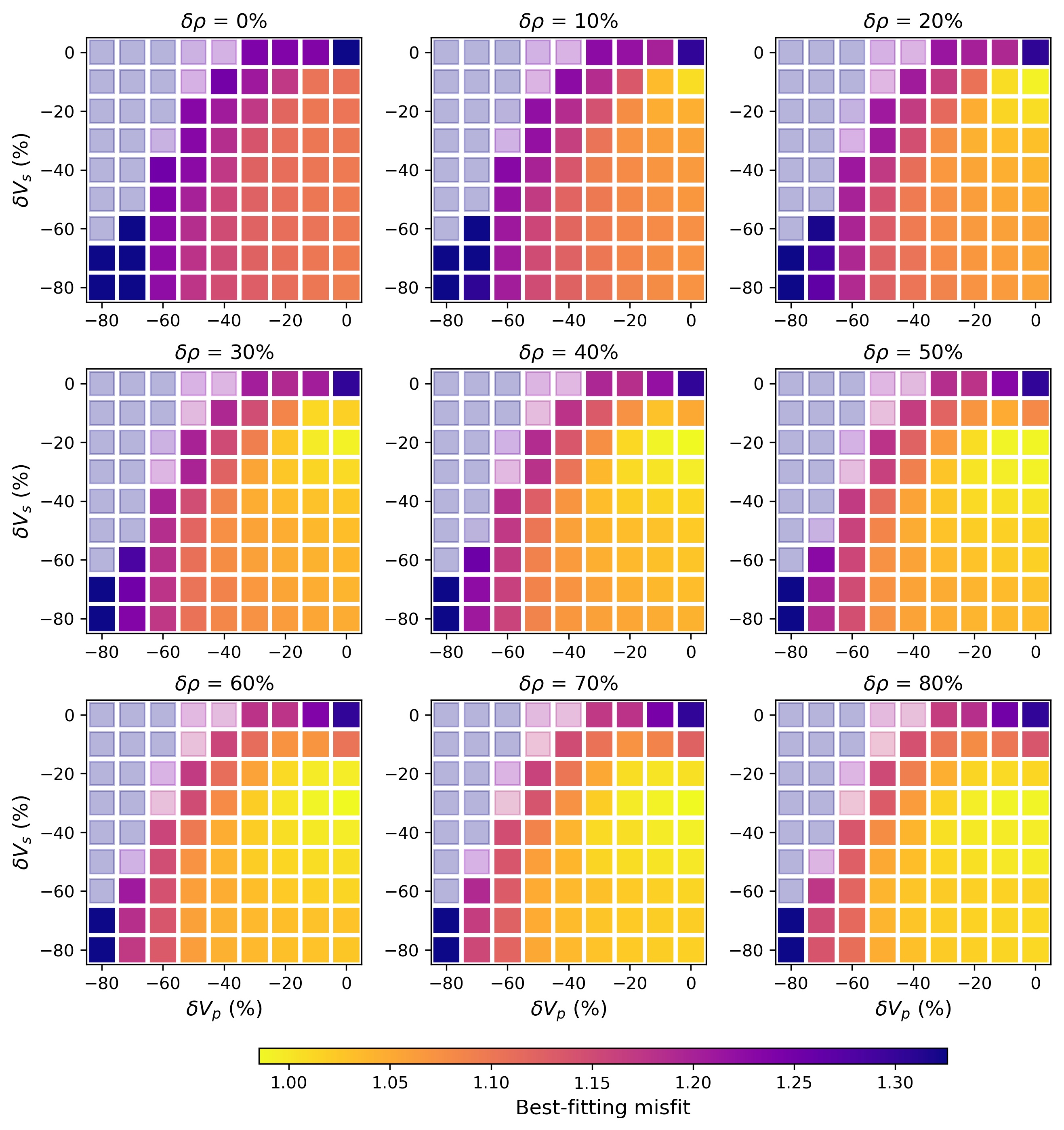}
    \caption{The best-fitting misfit as a function of $\delta V_{p}$ and $\delta V_{s}$, for constant values of $\delta \rho$. Percentage deviations are relative to the lowermost mantle in PREM. Squares with high transparency represent models that are excluded as they have negative bulk modulus.}
    \label{constant_rho}
\end{figure}

\begin{figure}[H]
    \centering
    \includegraphics[width=16cm]{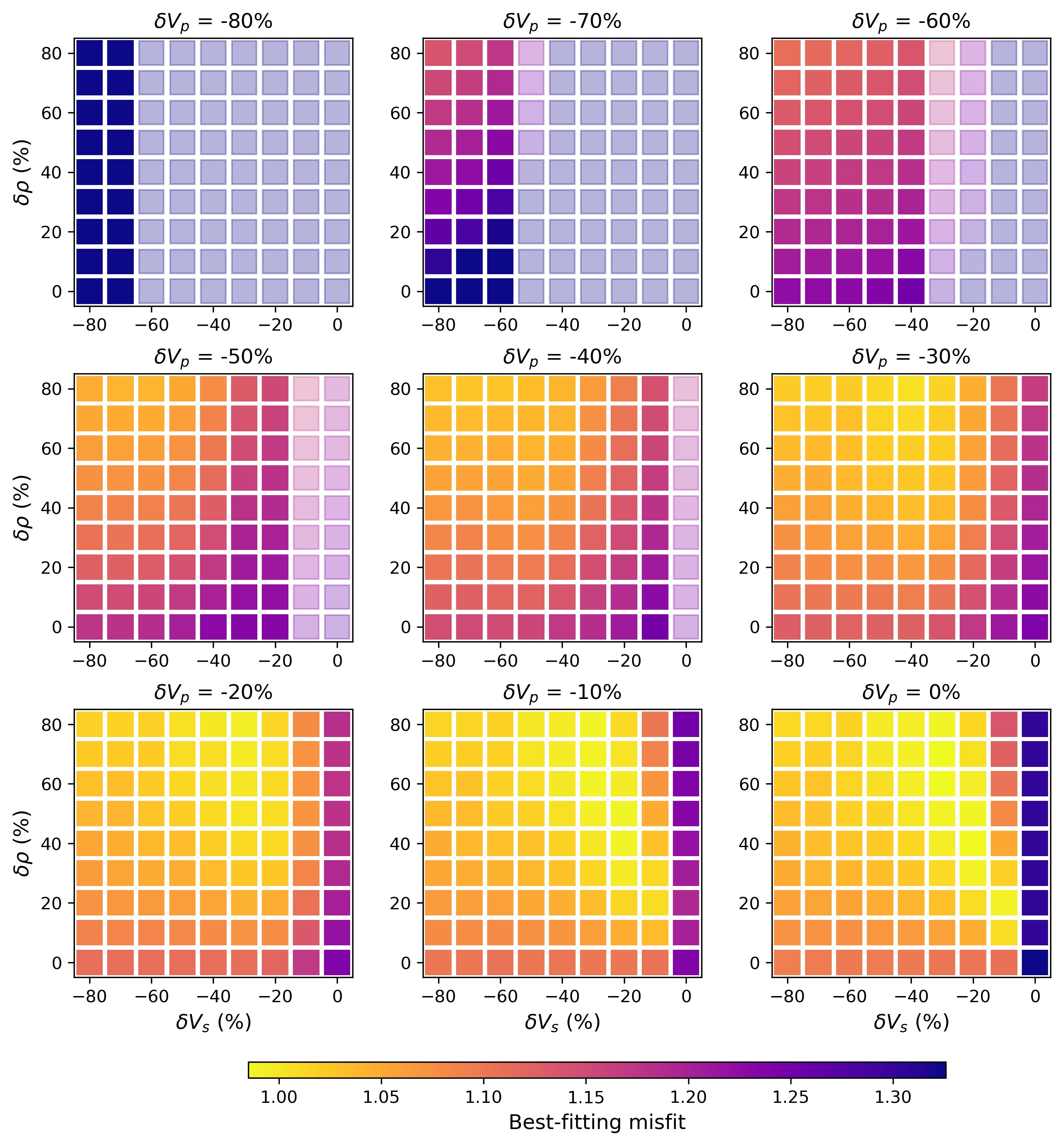}
    \caption{The best-fitting misfit as a function of $\delta V_{s}$ and $\delta \rho$, for constant values of $\delta V_{p}$. Percentage deviations are relative to the lowermost mantle in PREM. Squares with high transparency represent models that are excluded as they have negative bulk modulus.}
    \label{constant_vp}
\end{figure}

\begin{figure}[H]
    \centering
    \includegraphics[width=16cm]{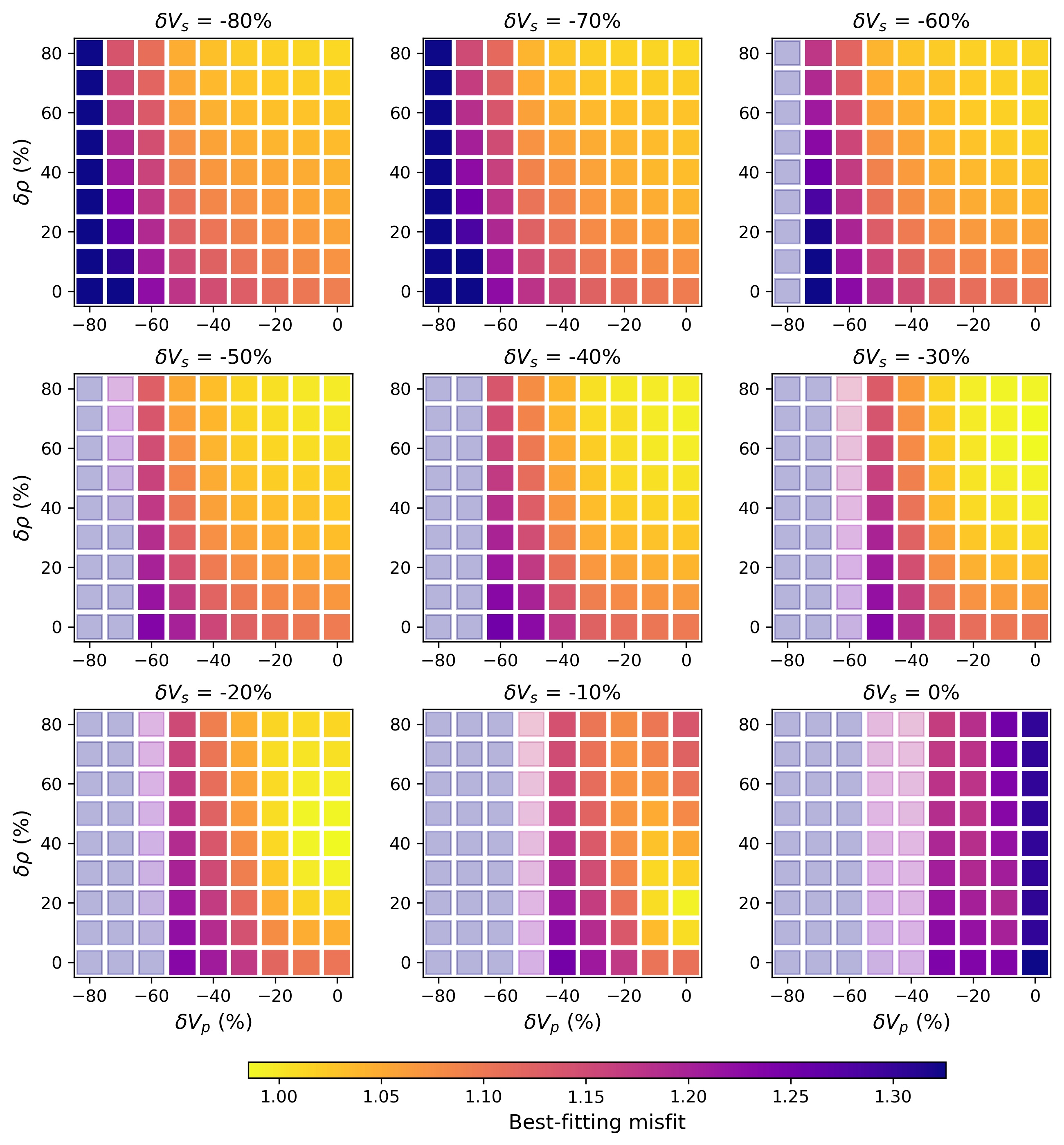}
    \caption{The best-fitting misfit as a function of $\delta V_{p}$ and $\delta \rho$, for constant values of $\delta V_{s}$. Percentage deviations are relative to the lowermost mantle in PREM. Squares with high transparency represent models that are excluded as they have negative bulk modulus.}
    \label{constant_vs}
\end{figure}

Figures \ref{constant_rho}, \ref{constant_vp}, and \ref{constant_vs} show the results for the grid search procedure in greater detail. Each figure shows slices through the misfit volume where Figures \ref{constant_rho}, \ref{constant_vp}, and \ref{constant_vs} are slices of constant P-velocity deviation, $\delta V_{p}$; S-velocity deviation, $\delta V_{p}$; and density deviation, $\delta \rho$, respectively - three sets of orthogonal slices through the same volume. All 729 models tested are presented and the 180 models that are excluded due to having negative bulk modules are indicated. These excluded models inhabit a poorly-fitting region of the parameter space with high $\delta V_{p}$ and low $\delta V_{s}$. These results highlight the very wide parameter space that is better-fitting than no layer, making a precise assessment of layer properties impossible.

We also perform a grid search for a totally molten layer with $\delta V_{s}$ fixed at -100\% and the same intervals of $\delta V_{s}$ and $\delta \rho$. None of the 81 models had a non-zero best-fitting thickness, reinforcing that a molten layer is not preferred over EPOC with no layer.

\pagebreak

\section{Mode dataset} \label{dataset}

Table \ref{dataset_table0} contains details of all 353 modes (334 spheroidal, 19 toroidal) that are included in the dataset. The centre frequencies and measurement errors are taken from the most recent study in which mode was measured \citep{Resovsky1998, Deuss2013, Koelemeijer2013, Schneider2021, REM}. Note that different studies take different approaches to defining the measurement error. For the solid layer used in the main paper, the rate of change of centre frequency with layer thickness for each mode is also given in the gradient column. The measured centre frequencies are corrected for the first-order effects of ellipticity \citep{Dahlen1999}, although this correction is minor and affects the best-fitting thickness of the solid layer by only 50~m.

\begin{figure}[H]
    \centering
    \includegraphics[width=16cm]{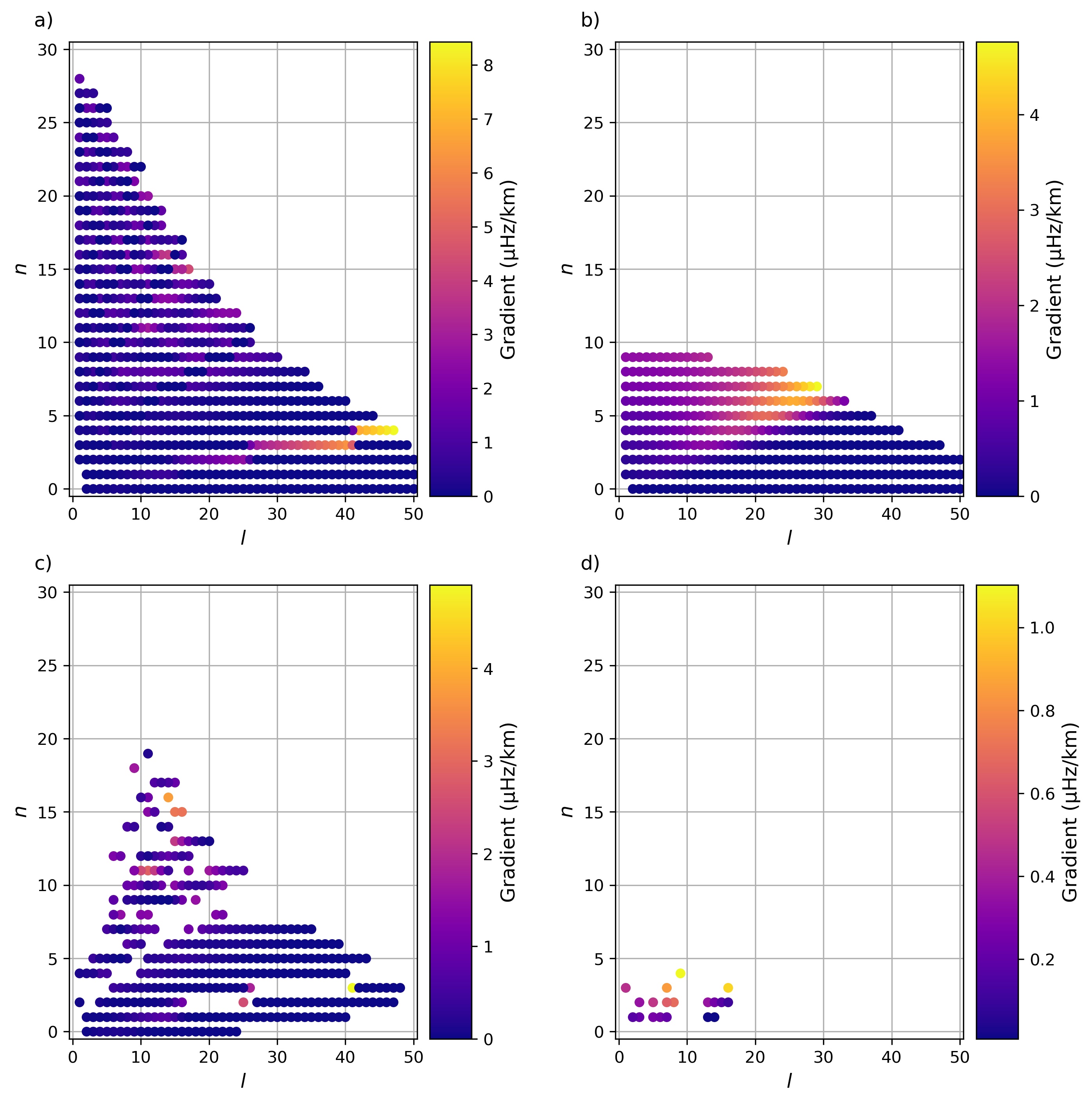}
    \caption{Absolute rate of change of centre frequency with increasing solid layer thickness plotted against $n$ and $l$ for (a) spheroidal modes from MINEOS, (b) toroidal modes from MINEOS, (c) the 334 spheroidal modes that are in the dataset and (d) the 19 toroidal modes that are in the dataset. Note the different colour scales for each axis.}
    \label{gradient}
\end{figure}

Figure~\ref{gradient} shows the absolute rate of change of the centre frequency for different modes for the solid layer. MINEOS \citep{Masters2011} calculates centre frequencies for all possible modes, however not all modes have been measured or are possible to observe. Figure~\ref{gradient}a shows all spheroidal modes outputted by MINEOS with centre frequencies up to 10~mHz. The spheroidal modes with the greatest sensitivity are concentrated in a diagonal band across all $l$ where $n$ is less than 5. This band includes the strongly CMB sensitive Stoneley modes, however many of these modes at higher $l$ values have not been measured. Figure~\ref{gradient}b is the same but for toroidal modes with centre frequencies up to 10~mHz. The 353 modes of the dataset are shown in Figure~\ref{gradient}c and \ref{gradient}d for spheroidal and toroidal modes, respectively.

\begin{table}[H]
	\centering
	\begin{tabular}{|l|c|c|c|c|c|}
		\hline
		Mode & \makecell{Centre frequency \\ (mHz)} & \makecell{Measurement \\ error, $\epsilon_{d}$ \\(\textmu Hz)} & \makecell{Model error, \\ $\epsilon_{f}$ (\textmu Hz)} & \makecell{Gradient \\ (\textmu Hz/km)} & Source \\ 
		\hline
		$_{0}$S$_{2}$ &  0.30968 &     0.02 &     0.28 &  0.03333 & D13 \\ 
		\hline
		$_{0}$S$_{3}$ &  0.46858 &     0.04 &     0.22 &  0.06759 & D13 \\ 
		\hline
		$_{0}$S$_{4}$ &  0.64685 &     0.03 &     0.07 &  0.10093 & D13 \\ 
		\hline
		$_{0}$S$_{5}$ &  0.84003 &     0.04 &     0.28 &  0.12679 & D13 \\ 
		\hline
		$_{0}$S$_{6}$ &  1.03756 &     0.04 &     0.44 &  0.14113 & D13 \\ 
		\hline
		$_{0}$S$_{7}$ &  1.23098 &     0.03 &     0.50 &  0.13806 & D13 \\ 
		\hline
		$_{0}$S$_{8}$ &  1.41279 &     0.02 &     0.45 &  0.11379 & D13 \\ 
		\hline
		$_{0}$S$_{9}$ &  1.57753 &     0.02 &     0.68 &  0.07650 & D13 \\ 
		\hline
		$_{0}$S$_{10}$ &  1.72561 &     0.11 &     1.02 &  0.04357 & R\&R98 \\ 
		\hline
		$_{0}$S$_{11}$ &  1.86185 &     0.08 &     1.36 &  0.02293 & D13 \\ 
		\hline
		$_{0}$S$_{12}$ &  1.98968 &     0.03 &     1.71 &  0.01214 & D13 \\ 
		\hline
		$_{0}$S$_{13}$ &  2.11197 &     0.04 &     2.07 &  0.00692 & D13 \\ 
		\hline
		$_{0}$S$_{14}$ &  2.23042 &     0.04 &     2.43 &  0.00447 & D13 \\ 
		\hline
		$_{0}$S$_{15}$ &  2.34541 &     0.06 &     2.79 &  0.00330 & D13 \\ 
		\hline
		$_{0}$S$_{16}$ &  2.45746 &     0.03 &     3.13 &  0.00272 & D13 \\ 
		\hline
		$_{0}$S$_{17}$ &  2.56649 &     0.05 &     3.46 &  0.00240 & D13 \\ 
		\hline
		$_{0}$S$_{18}$ &  2.67242 &     0.12 &     3.77 &  0.00219 & R\&R98 \\ 
		\hline
		$_{0}$S$_{19}$ &  2.77682 &     0.10 &     4.03 &  0.00203 & D13 \\ 
		\hline
		$_{0}$S$_{20}$ &  2.87833 &     0.16 &     4.26 &  0.00190 & D13 \\ 
		\hline
		$_{0}$S$_{21}$ &  2.97745 &     0.61 &     4.45 &  0.00178 & D13 \\ 
		\hline
		$_{0}$S$_{22}$ &  3.07457 &     0.15 &     4.60 &  0.00168 & M\&W95 \\ 
		\hline
		$_{0}$S$_{23}$ &  3.17062 &     0.10 &     4.70 &  0.00157 & M\&W95 \\ 
		\hline
		$_{0}$S$_{24}$ &  3.26558 &     0.10 &     4.76 &  0.00148 & M\&W95 \\ 
		\hline
		$_{1}$S$_{2}$ &  0.67994 &     0.05 &     0.24 &  0.03920 & D13 \\ 
		\hline
		$_{1}$S$_{3}$ &  0.94005 &     0.06 &     0.06 &  0.02457 & D13 \\ 
		\hline
		$_{1}$S$_{4}$ &  1.17300 &     0.05 &     0.14 &  0.01992 & D13 \\ 
		\hline
		$_{1}$S$_{5}$ &  1.37022 &     0.03 &     1.05 &  0.03762 & D13 \\ 
		\hline
		$_{1}$S$_{6}$ &  1.52159 &     0.04 &     2.29 &  0.08397 & D13 \\ 
		\hline
		$_{1}$S$_{7}$ &  1.65466 &     0.06 &     2.89 &  0.14636 & D13 \\ 
		\hline
		$_{1}$S$_{8}$ &  1.79795 &     0.03 &     3.45 &  0.23406 & D13 \\ 
		\hline
		$_{1}$S$_{9}$ &  1.96204 &     0.03 &     4.12 &  0.34118 & D13 \\ 
		\hline
		$_{1}$S$_{10}$ &  2.14640 &     0.14 &     4.81 &  0.44961 & D13 \\ 
		\hline
		$_{1}$S$_{11}$ &  2.34574 &     0.41 &     5.44 &  0.54974 & K13 \\ 
		\hline
		$_{1}$S$_{12}$ &  2.55265 &     0.09 &     5.95 &  0.63958 & K13 \\ 
		\hline
		$_{1}$S$_{13}$ &  2.76442 &     0.11 &     6.16 &  0.71062 & K13 \\ 
		\hline
		$_{1}$S$_{14}$ &  2.97383 &     0.15 &     5.20 &  0.71055 & K13 \\ 
		\hline
		$_{1}$S$_{15}$ &  3.16906 &     0.10 &     2.41 &  0.43857 & K13 \\ 
		\hline
		$_{1}$S$_{16}$ &  3.33768 &     0.08 &     2.89 &  0.11158 & K13 \\ 
		\hline
		$_{1}$S$_{17}$ &  3.49348 &     0.84 &     3.12 &  0.02835 & REM \\ 
		\hline
		$_{1}$S$_{18}$ &  3.64383 &     0.61 &     3.15 &  0.00828 & REM \\ 
		\hline
	\end{tabular}
	\caption{Properties and composition of the modes in the dataset. Modes are ordered by $n$ and then $l$ with spheroidal modes followed by toroidal modes. Gradients given are for the solid layer. Sources: REM \citep{REM}, M\&W95 \citep{Masters1995}, R\&R98 \citep{Resovsky1998}, D13 \citep{Deuss2013}, K13 \citep{Koelemeijer2013}, S\&D21 \citep{Schneider2021}.}
	\label{dataset_table0}
\end{table}

\pagebreak
\setcounter{table}{0}

\begin{table}[H]
	\centering
	\begin{tabular}{|l|c|c|c|c|c|}
		\hline
		Mode & \makecell{Centre frequency \\ (mHz)} & \makecell{Measurement \\ error, $\epsilon_{d}$ \\(\textmu Hz)} & \makecell{Model error, \\ $\epsilon_{f}$ (\textmu Hz)} & \makecell{Gradient \\ (\textmu Hz/km)} & Source \\ 
		\hline
		$_{1}$S$_{19}$ &  3.79231 &     0.91 &     3.06 &  0.00209 & REM \\ 
		\hline
		$_{1}$S$_{20}$ &  3.94004 &     0.76 &     2.92 &  0.00014 & REM \\ 
		\hline
		$_{1}$S$_{21}$ &  4.08571 &     0.61 &     2.78 &  0.00087 & REM \\ 
		\hline
		$_{1}$S$_{22}$ &  4.23153 &     0.80 &     2.65 &  0.00110 & REM \\ 
		\hline
		$_{1}$S$_{23}$ &  4.37630 &     1.16 &     2.54 &  0.00110 & REM \\ 
		\hline
		$_{1}$S$_{24}$ &  4.52108 &     1.16 &     2.46 &  0.00102 & REM \\ 
		\hline
		$_{1}$S$_{25}$ &  4.66269 &     2.10 &     2.42 &  0.00090 & REM \\ 
		\hline
		$_{1}$S$_{26}$ &  4.80915 &     1.38 &     2.42 &  0.00076 & REM \\ 
		\hline
		$_{1}$S$_{27}$ &  4.95281 &     1.59 &     2.47 &  0.00064 & REM \\ 
		\hline
		$_{1}$S$_{28}$ &  5.08868 &     2.56 &     2.56 &  0.00053 & REM \\ 
		\hline
		$_{1}$S$_{29}$ &  5.23305 &     1.53 &     2.71 &  0.00040 & REM \\ 
		\hline
		$_{1}$S$_{30}$ &  5.37308 &     1.78 &     2.91 &  0.00031 & REM \\ 
		\hline
		$_{1}$S$_{31}$ &  5.51356 &     2.08 &     3.17 &  0.00022 & REM \\ 
		\hline
		$_{1}$S$_{32}$ &  5.65238 &     2.33 &     3.48 &  0.00013 & REM \\ 
		\hline
		$_{1}$S$_{33}$ &  5.78883 &     2.06 &     3.85 &  0.00003 & REM \\ 
		\hline
		$_{1}$S$_{34}$ &  5.92921 &     2.90 &     4.28 &  0.00015 & REM \\ 
		\hline
		$_{1}$S$_{35}$ &  6.06184 &     2.37 &     4.77 &  0.00009 & REM \\ 
		\hline
		$_{1}$S$_{36}$ &  6.19501 &     3.00 &     5.31 &  0.00015 & REM \\ 
		\hline
		$_{1}$S$_{37}$ &  6.33199 &     3.30 &     5.90 &  0.00015 & REM \\ 
		\hline
		$_{1}$S$_{38}$ &  6.46535 &     2.65 &     6.54 &  0.00023 & REM \\ 
		\hline
		$_{1}$S$_{39}$ &  6.59433 &     3.40 &     7.22 &  0.00028 & REM \\ 
		\hline
		$_{1}$S$_{40}$ &  6.72829 &     3.85 &     7.95 &  0.00032 & REM \\ 
		\hline
		$_{2}$S$_{1}$ &  0.40418 &     0.04 &     0.66 &  0.05505 & D13 \\ 
		\hline
		$_{2}$S$_{4}$ &  1.37950 &     0.03 &     2.52 &  0.10600 & D13 \\ 
		\hline
		$_{2}$S$_{5}$ &  1.51522 &     0.03 &     1.84 &  0.06976 & D13 \\ 
		\hline
		$_{2}$S$_{6}$ &  1.68113 &     0.06 &     1.82 &  0.03611 & D13 \\ 
		\hline
		$_{2}$S$_{7}$ &  1.86517 &     0.05 &     2.16 &  0.02241 & D13 \\ 
		\hline
		$_{2}$S$_{8}$ &  2.04948 &     0.03 &     2.44 &  0.01879 & D13 \\ 
		\hline
		$_{2}$S$_{9}$ &  2.22867 &     0.15 &     2.70 &  0.01953 & D13 \\ 
		\hline
		$_{2}$S$_{10}$ &  2.40320 &     0.01 &     2.92 &  0.02348 & D13 \\ 
		\hline
		$_{2}$S$_{11}$ &  2.57235 &     0.18 &     3.08 &  0.03175 & D13 \\ 
		\hline
		$_{2}$S$_{12}$ &  2.73726 &     0.02 &     3.13 &  0.04822 & D13 \\ 
		\hline
		$_{2}$S$_{13}$ &  2.89989 &     0.04 &     3.00 &  0.08586 & D13 \\ 
		\hline
		$_{2}$S$_{14}$ &  3.06236 &     0.07 &     2.30 &  0.19969 & K13 \\ 
		\hline
		$_{2}$S$_{15}$ &  3.23880 &     0.03 &     4.90 &  0.59212 & K13 \\ 
		\hline
		$_{2}$S$_{16}$ &  3.44091 &     0.14 &     8.90 &  1.04683 & K13 \\ 
		\hline
		$_{2}$S$_{25}$ &  5.39732 &     0.20 &    15.52 &  2.55813 & K13 \\ 
		\hline
		$_{2}$S$_{27}$ &  5.74614 &     0.85 &     4.08 &  0.02412 & REM \\ 
		\hline
		$_{2}$S$_{28}$ &  5.90383 &     1.10 &     4.32 &  0.00368 & REM \\ 
		\hline
		$_{2}$S$_{29}$ &  6.06816 &     1.28 &     4.56 &  0.00077 & REM \\ 
		\hline
	\end{tabular}
	\caption{Continued: properties and composition of the modes in the dataset. Modes are ordered by $n$ and then $l$ with spheroidal modes followed by toroidal modes. Gradients given are for the solid layer. Sources: REM \citep{REM}, M\&W95 \citep{Masters1995}, R\&R98 \citep{Resovsky1998}, D13 \citep{Deuss2013}, K13 \citep{Koelemeijer2013}, S\&D21 \citep{Schneider2021}.}
	\label{dataset_table1}
\end{table}

\pagebreak
\setcounter{table}{0}

\begin{table}[H]
	\centering
	\begin{tabular}{|l|c|c|c|c|c|}
		\hline
		Mode & \makecell{Centre frequency \\ (mHz)} & \makecell{Measurement \\ error, $\epsilon_{d}$ \\(\textmu Hz)} & \makecell{Model error, \\ $\epsilon_{f}$ (\textmu Hz)} & \makecell{Gradient \\ (\textmu Hz/km)} & Source \\ 
		\hline
		$_{2}$S$_{30}$ &  6.22858 &     1.17 &     4.79 &  0.00000 & REM \\ 
		\hline
		$_{2}$S$_{31}$ &  6.38518 &     1.13 &     5.02 &  0.00025 & REM \\ 
		\hline
		$_{2}$S$_{32}$ &  6.54111 &     1.49 &     5.25 &  0.00036 & REM \\ 
		\hline
		$_{2}$S$_{33}$ &  6.69721 &     1.34 &     5.48 &  0.00042 & REM \\ 
		\hline
		$_{2}$S$_{34}$ &  6.85210 &     1.25 &     5.70 &  0.00045 & REM \\ 
		\hline
		$_{2}$S$_{35}$ &  7.01193 &     2.31 &     5.91 &  0.00046 & REM \\ 
		\hline
		$_{2}$S$_{36}$ &  7.16443 &     2.45 &     6.12 &  0.00047 & REM \\ 
		\hline
		$_{2}$S$_{37}$ &  7.31857 &     2.07 &     6.31 &  0.00047 & REM \\ 
		\hline
		$_{2}$S$_{38}$ &  7.47302 &     1.81 &     6.49 &  0.00049 & REM \\ 
		\hline
		$_{2}$S$_{39}$ &  7.62341 &     2.66 &     6.64 &  0.00046 & REM \\ 
		\hline
		$_{2}$S$_{40}$ &  7.77496 &     2.50 &     6.78 &  0.00046 & REM \\ 
		\hline
		$_{2}$S$_{41}$ &  7.92126 &     3.89 &     6.88 &  0.00044 & REM \\ 
		\hline
		$_{2}$S$_{42}$ &  8.06990 &     4.58 &     6.96 &  0.00041 & REM \\ 
		\hline
		$_{2}$S$_{43}$ &  8.21956 &     4.06 &     7.01 &  0.00040 & REM \\ 
		\hline
		$_{2}$S$_{44}$ &  8.36056 &     4.24 &     7.02 &  0.00038 & REM \\ 
		\hline
		$_{2}$S$_{45}$ &  8.50898 &     4.44 &     7.00 &  0.00034 & REM \\ 
		\hline
		$_{2}$S$_{46}$ &  8.65704 &     4.73 &     6.95 &  0.00032 & REM \\ 
		\hline
		$_{2}$S$_{47}$ &  8.80729 &     5.10 &     6.86 &  0.00028 & REM \\ 
		\hline
		$_{3}$S$_{6}$ &  2.54880 &     0.08 &     2.33 &  0.43663 & D13 \\ 
		\hline
		$_{3}$S$_{7}$ &  2.68577 &     0.21 &     2.22 &  0.35142 & D13 \\ 
		\hline
		$_{3}$S$_{8}$ &  2.81924 &     0.03 &     2.02 &  0.26385 & D13 \\ 
		\hline
		$_{3}$S$_{9}$ &  2.95137 &     0.03 &     1.76 &  0.19130 & D13 \\ 
		\hline
		$_{3}$S$_{10}$ &  3.08210 &     0.50 &     1.51 &  0.13793 & REM \\ 
		\hline
		$_{3}$S$_{11}$ &  3.21950 &     0.53 &     1.27 &  0.10097 & REM \\ 
		\hline
		$_{3}$S$_{12}$ &  3.36134 &     0.41 &     1.06 &  0.07605 & REM \\ 
		\hline
		$_{3}$S$_{13}$ &  3.50753 &     0.55 &     1.15 &  0.05938 & REM \\ 
		\hline
		$_{3}$S$_{14}$ &  3.65618 &     0.53 &     1.35 &  0.04793 & REM \\ 
		\hline
		$_{3}$S$_{15}$ &  3.81096 &     0.59 &     1.56 &  0.03975 & REM \\ 
		\hline
		$_{3}$S$_{16}$ &  3.96683 &     0.65 &     1.76 &  0.03368 & REM \\ 
		\hline
		$_{3}$S$_{17}$ &  4.12400 &     0.60 &     1.97 &  0.02899 & REM \\ 
		\hline
		$_{3}$S$_{18}$ &  4.28379 &     0.67 &     2.17 &  0.02534 & REM \\ 
		\hline
		$_{3}$S$_{19}$ &  4.44612 &     0.67 &     2.38 &  0.02255 & REM \\ 
		\hline
		$_{3}$S$_{20}$ &  4.60897 &     0.88 &     2.59 &  0.02058 & REM \\ 
		\hline
		$_{3}$S$_{21}$ &  4.77158 &     0.75 &     2.79 &  0.01955 & REM \\ 
		\hline
		$_{3}$S$_{22}$ &  4.93287 &     0.85 &     3.00 &  0.01987 & REM \\ 
		\hline
		$_{3}$S$_{23}$ &  5.09842 &     0.76 &     3.19 &  0.02270 & REM \\ 
		\hline
		$_{3}$S$_{24}$ &  5.26295 &     0.81 &     3.34 &  0.03278 & REM \\ 
		\hline
		$_{3}$S$_{25}$ &  5.42711 &     0.15 &     2.73 &  0.08664 & K13 \\ 
		\hline
		$_{3}$S$_{26}$ &  5.62085 &     0.24 &    17.54 &  1.80030 & K13 \\ 
		\hline
		$_{3}$S$_{41}$ &  8.82312 &     2.57 &     1.82 &  4.90257 & REM \\ 
		\hline
	\end{tabular}
	\caption{Continued: properties and composition of the modes in the dataset. Modes are ordered by $n$ and then $l$ with spheroidal modes followed by toroidal modes. Gradients given are for the solid layer. Sources: REM \citep{REM}, M\&W95 \citep{Masters1995}, R\&R98 \citep{Resovsky1998}, D13 \citep{Deuss2013}, K13 \citep{Koelemeijer2013}, S\&D21 \citep{Schneider2021}.}
	\label{dataset_table2}
\end{table}

\pagebreak
\setcounter{table}{0}

\begin{table}[H]
	\centering
	\begin{tabular}{|l|c|c|c|c|c|}
		\hline
		Mode & \makecell{Centre frequency \\ (mHz)} & \makecell{Measurement \\ error, $\epsilon_{d}$ \\(\textmu Hz)} & \makecell{Model error, \\ $\epsilon_{f}$ (\textmu Hz)} & \makecell{Gradient \\ (\textmu Hz/km)} & Source \\ 
		\hline
		$_{3}$S$_{42}$ &  8.97689 &     2.41 &     1.36 &  0.00091 & REM \\ 
		\hline
		$_{3}$S$_{43}$ &  9.13826 &     2.30 &     0.87 &  0.00021 & REM \\ 
		\hline
		$_{3}$S$_{44}$ &  9.29013 &     2.43 &     0.34 &  0.00002 & REM \\ 
		\hline
		$_{3}$S$_{45}$ &  9.44223 &     3.31 &     0.23 &  0.00015 & REM \\ 
		\hline
		$_{3}$S$_{46}$ &  9.60346 &     2.47 &     0.83 &  0.00000 & REM \\ 
		\hline
		$_{3}$S$_{47}$ &  9.75065 &     2.46 &     1.48 &  0.00000 & REM \\ 
		\hline
		$_{3}$S$_{48}$ &  9.90813 &     3.93 &     2.16 &  0.00011 & REM \\ 
		\hline
		$_{4}$S$_{1}$ &  1.41179 &     0.05 &     0.40 &  0.13150 & D13 \\ 
		\hline
		$_{4}$S$_{2}$ &  1.72140 &     0.05 &     0.51 &  0.14143 & D13 \\ 
		\hline
		$_{4}$S$_{3}$ &  2.04827 &     0.01 &     0.39 &  0.16655 & D13 \\ 
		\hline
		$_{4}$S$_{4}$ &  2.27830 &     0.03 &     1.65 &  0.49031 & D13 \\ 
		\hline
		$_{4}$S$_{5}$ &  2.41112 &     0.03 &     2.17 &  0.49078 & D13 \\ 
		\hline
		$_{4}$S$_{10}$ &  3.86407 &     0.63 &     2.40 &  0.41341 & REM \\ 
		\hline
		$_{4}$S$_{11}$ &  4.00750 &     1.00 &     2.55 &  0.38606 & REM \\ 
		\hline
		$_{4}$S$_{12}$ &  4.15199 &     1.50 &     2.56 &  0.32587 & REM \\ 
		\hline
		$_{4}$S$_{13}$ &  4.29204 &     1.04 &     2.51 &  0.26169 & REM \\ 
		\hline
		$_{4}$S$_{14}$ &  4.43529 &     0.81 &     2.47 &  0.20459 & REM \\ 
		\hline
		$_{4}$S$_{15}$ &  4.58399 &     1.50 &     2.44 &  0.15788 & REM \\ 
		\hline
		$_{4}$S$_{16}$ &  4.72984 &     1.50 &     2.44 &  0.12105 & REM \\ 
		\hline
		$_{4}$S$_{17}$ &  4.88531 &     1.14 &     2.46 &  0.09255 & REM \\ 
		\hline
		$_{4}$S$_{18}$ &  5.04365 &     1.05 &     2.51 &  0.07067 & REM \\ 
		\hline
		$_{4}$S$_{19}$ &  5.20062 &     1.53 &     2.58 &  0.05392 & REM \\ 
		\hline
		$_{4}$S$_{20}$ &  5.36218 &     1.26 &     2.67 &  0.04110 & REM \\ 
		\hline
		$_{4}$S$_{21}$ &  5.52606 &     1.29 &     2.78 &  0.03127 & REM \\ 
		\hline
		$_{4}$S$_{22}$ &  5.69499 &     1.50 &     2.90 &  0.02374 & REM \\ 
		\hline
		$_{4}$S$_{23}$ &  5.86147 &     1.34 &     3.03 &  0.01798 & REM \\ 
		\hline
		$_{4}$S$_{24}$ &  6.02866 &     1.49 &     3.16 &  0.01360 & REM \\ 
		\hline
		$_{4}$S$_{25}$ &  6.19722 &     1.58 &     3.30 &  0.01027 & REM \\ 
		\hline
		$_{4}$S$_{26}$ &  6.36548 &     1.24 &     3.43 &  0.00775 & REM \\ 
		\hline
		$_{4}$S$_{27}$ &  6.53553 &     1.73 &     3.56 &  0.00589 & REM \\ 
		\hline
		$_{4}$S$_{28}$ &  6.70264 &     1.86 &     3.67 &  0.00447 & REM \\ 
		\hline
		$_{4}$S$_{29}$ &  6.87294 &     1.30 &     3.76 &  0.00343 & REM \\ 
		\hline
		$_{4}$S$_{30}$ &  7.03810 &     1.30 &     3.83 &  0.00266 & REM \\ 
		\hline
		$_{4}$S$_{31}$ &  7.20440 &     2.12 &     3.86 &  0.00209 & REM \\ 
		\hline
		$_{4}$S$_{32}$ &  7.36938 &     1.58 &     3.87 &  0.00167 & REM \\ 
		\hline
		$_{4}$S$_{33}$ &  7.53650 &     1.87 &     3.83 &  0.00136 & REM \\ 
		\hline
		$_{4}$S$_{34}$ &  7.70007 &     1.83 &     3.74 &  0.00113 & REM \\ 
		\hline
		$_{4}$S$_{35}$ &  7.85957 &     1.78 &     3.61 &  0.00096 & REM \\ 
		\hline
		$_{4}$S$_{36}$ &  8.01994 &     1.48 &     3.43 &  0.00083 & REM \\ 
		\hline
		$_{4}$S$_{37}$ &  8.18437 &     1.48 &     3.21 &  0.00074 & REM \\ 
		\hline
	\end{tabular}
	\caption{Continued: properties and composition of the modes in the dataset. Modes are ordered by $n$ and then $l$ with spheroidal modes followed by toroidal modes. Gradients given are for the solid layer. Sources: REM \citep{REM}, M\&W95 \citep{Masters1995}, R\&R98 \citep{Resovsky1998}, D13 \citep{Deuss2013}, K13 \citep{Koelemeijer2013}, S\&D21 \citep{Schneider2021}.}
	\label{dataset_table3}
\end{table}

\pagebreak
\setcounter{table}{0}

\begin{table}[H]
	\centering
	\begin{tabular}{|l|c|c|c|c|c|}
		\hline
		Mode & \makecell{Centre frequency \\ (mHz)} & \makecell{Measurement \\ error, $\epsilon_{d}$ \\(\textmu Hz)} & \makecell{Model error, \\ $\epsilon_{f}$ (\textmu Hz)} & \makecell{Gradient \\ (\textmu Hz/km)} & Source \\ 
		\hline
		$_{4}$S$_{38}$ &  8.34214 &     2.15 &     2.93 &  0.00069 & REM \\ 
		\hline
		$_{4}$S$_{39}$ &  8.49951 &     2.34 &     2.60 &  0.00072 & REM \\ 
		\hline
		$_{4}$S$_{40}$ &  8.66348 &     2.81 &     2.28 &  0.00110 & REM \\ 
		\hline
		$_{5}$S$_{3}$ &  2.16868 &     0.06 &     0.98 &  0.39228 & D13 \\ 
		\hline
		$_{5}$S$_{4}$ &  2.37918 &     0.04 &     0.30 &  0.14146 & D13 \\ 
		\hline
		$_{5}$S$_{5}$ &  2.70340 &     0.01 &     0.27 &  0.11232 & D13 \\ 
		\hline
		$_{5}$S$_{6}$ &  3.01105 &     0.05 &     0.43 &  0.04457 & D13 \\ 
		\hline
		$_{5}$S$_{7}$ &  3.29165 &     0.04 &     0.53 &  0.00336 & D13 \\ 
		\hline
		$_{5}$S$_{8}$ &  3.52593 &     0.02 &     1.09 &  0.12761 & D13 \\ 
		\hline
		$_{5}$S$_{11}$ &  4.45688 &     0.11 &     0.44 &  0.24028 & D13 \\ 
		\hline
		$_{5}$S$_{12}$ &  4.69577 &     0.03 &     0.49 &  0.22440 & D13 \\ 
		\hline
		$_{5}$S$_{13}$ &  4.92550 &     1.02 &     0.65 &  0.24944 & REM \\ 
		\hline
		$_{5}$S$_{14}$ &  5.13496 &     0.06 &     1.60 &  0.30165 & D13 \\ 
		\hline
		$_{5}$S$_{15}$ &  5.32687 &     0.02 &     2.75 &  0.34504 & D13 \\ 
		\hline
		$_{5}$S$_{16}$ &  5.50245 &     0.10 &     3.64 &  0.34787 & D13 \\ 
		\hline
		$_{5}$S$_{17}$ &  5.66876 &     0.06 &     4.13 &  0.31599 & D13 \\ 
		\hline
		$_{5}$S$_{18}$ &  5.82920 &     1.96 &     4.35 &  0.26906 & REM \\ 
		\hline
		$_{5}$S$_{19}$ &  5.98849 &     2.17 &     4.42 &  0.22024 & REM \\ 
		\hline
		$_{5}$S$_{20}$ &  6.15222 &     1.88 &     4.39 &  0.17555 & REM \\ 
		\hline
		$_{5}$S$_{21}$ &  6.31038 &     2.18 &     4.30 &  0.13712 & REM \\ 
		\hline
		$_{5}$S$_{22}$ &  6.47356 &     1.85 &     4.15 &  0.10525 & REM \\ 
		\hline
		$_{5}$S$_{23}$ &  6.63542 &     1.79 &     3.94 &  0.07951 & REM \\ 
		\hline
		$_{5}$S$_{24}$ &  6.80077 &     2.32 &     3.66 &  0.05914 & REM \\ 
		\hline
		$_{5}$S$_{25}$ &  6.96593 &     1.92 &     3.30 &  0.04333 & REM \\ 
		\hline
		$_{5}$S$_{26}$ &  7.13265 &     1.79 &     2.86 &  0.03130 & REM \\ 
		\hline
		$_{5}$S$_{27}$ &  7.29192 &     2.22 &     2.31 &  0.02232 & REM \\ 
		\hline
		$_{5}$S$_{28}$ &  7.45535 &     2.08 &     1.67 &  0.01575 & REM \\ 
		\hline
		$_{5}$S$_{29}$ &  7.61687 &     2.41 &     0.95 &  0.01102 & REM \\ 
		\hline
		$_{5}$S$_{30}$ &  7.77851 &     2.60 &     0.14 &  0.00769 & REM \\ 
		\hline
		$_{5}$S$_{31}$ &  7.94177 &     2.35 &     0.78 &  0.00537 & REM \\ 
		\hline
		$_{5}$S$_{32}$ &  8.09905 &     2.06 &     1.65 &  0.00377 & REM \\ 
		\hline
		$_{5}$S$_{33}$ &  8.25358 &     2.54 &     2.52 &  0.00268 & REM \\ 
		\hline
		$_{5}$S$_{34}$ &  8.40899 &     4.00 &     3.36 &  0.00195 & REM \\ 
		\hline
		$_{5}$S$_{35}$ &  8.57058 &     2.82 &     4.14 &  0.00145 & REM \\ 
		\hline
		$_{5}$S$_{36}$ &  8.72694 &     3.44 &     4.84 &  0.00111 & REM \\ 
		\hline
		$_{5}$S$_{37}$ &  8.88449 &     2.89 &     5.43 &  0.00089 & REM \\ 
		\hline
		$_{5}$S$_{38}$ &  9.04366 &     2.70 &     5.91 &  0.00074 & REM \\ 
		\hline
		$_{5}$S$_{39}$ &  9.20062 &     3.12 &     6.25 &  0.00062 & REM \\ 
		\hline
		$_{5}$S$_{40}$ &  9.36048 &     3.77 &     6.46 &  0.00055 & REM \\ 
		\hline
		$_{5}$S$_{41}$ &  9.50999 &     5.00 &     6.54 &  0.00048 & REM \\ 
		\hline
	\end{tabular}
	\caption{Continued: properties and composition of the modes in the dataset. Modes are ordered by $n$ and then $l$ with spheroidal modes followed by toroidal modes. Gradients given are for the solid layer. Sources: REM \citep{REM}, M\&W95 \citep{Masters1995}, R\&R98 \citep{Resovsky1998}, D13 \citep{Deuss2013}, K13 \citep{Koelemeijer2013}, S\&D21 \citep{Schneider2021}.}
	\label{dataset_table4}
\end{table}

\pagebreak
\setcounter{table}{0}

\begin{table}[H]
	\centering
	\begin{tabular}{|l|c|c|c|c|c|}
		\hline
		Mode & \makecell{Centre frequency \\ (mHz)} & \makecell{Measurement \\ error, $\epsilon_{d}$ \\(\textmu Hz)} & \makecell{Model error, \\ $\epsilon_{f}$ (\textmu Hz)} & \makecell{Gradient \\ (\textmu Hz/km)} & Source \\ 
		\hline
		$_{5}$S$_{42}$ &  9.67931 &     3.10 &     6.49 &  0.00046 & REM \\ 
		\hline
		$_{5}$S$_{43}$ &  9.83576 &     4.64 &     6.32 &  0.00042 & REM \\ 
		\hline
		$_{6}$S$_{8}$ &  3.73751 &     2.00 &     2.11 &  0.76092 & REM \\ 
		\hline
		$_{6}$S$_{9}$ &  3.96501 &     0.04 &     0.96 &  0.47702 & D13 \\ 
		\hline
		$_{6}$S$_{10}$ &  4.21105 &     0.03 &     0.53 &  0.31143 & D13 \\ 
		\hline
		$_{6}$S$_{14}$ &  5.41101 &     2.00 &     1.70 &  0.50727 & REM \\ 
		\hline
		$_{6}$S$_{15}$ &  5.60125 &     0.13 &     2.69 &  0.35179 & D13 \\ 
		\hline
		$_{6}$S$_{16}$ &  5.80677 &     1.33 &     4.01 &  0.24289 & REM \\ 
		\hline
		$_{6}$S$_{17}$ &  6.02074 &     1.13 &     5.13 &  0.18121 & REM \\ 
		\hline
		$_{6}$S$_{18}$ &  6.23570 &     0.08 &     6.09 &  0.14973 & D13 \\ 
		\hline
		$_{6}$S$_{19}$ &  6.44619 &     1.76 &     6.90 &  0.13545 & REM \\ 
		\hline
		$_{6}$S$_{20}$ &  6.65393 &     1.31 &     7.54 &  0.13072 & REM \\ 
		\hline
		$_{6}$S$_{21}$ &  6.85524 &     1.80 &     7.96 &  0.13039 & REM \\ 
		\hline
		$_{6}$S$_{22}$ &  7.05035 &     1.64 &     8.14 &  0.12996 & REM \\ 
		\hline
		$_{6}$S$_{23}$ &  7.23478 &     1.48 &     8.13 &  0.12588 & REM \\ 
		\hline
		$_{6}$S$_{24}$ &  7.41252 &     2.08 &     8.03 &  0.11666 & REM \\ 
		\hline
		$_{6}$S$_{25}$ &  7.58815 &     2.01 &     7.93 &  0.10305 & REM \\ 
		\hline
		$_{6}$S$_{26}$ &  7.75615 &     2.10 &     7.87 &  0.08710 & REM \\ 
		\hline
		$_{6}$S$_{27}$ &  7.92195 &     1.63 &     7.85 &  0.07089 & REM \\ 
		\hline
		$_{6}$S$_{28}$ &  8.08819 &     2.04 &     7.85 &  0.05593 & REM \\ 
		\hline
		$_{6}$S$_{29}$ &  8.25579 &     2.03 &     7.82 &  0.04297 & REM \\ 
		\hline
		$_{6}$S$_{30}$ &  8.41722 &     3.07 &     7.74 &  0.03225 & REM \\ 
		\hline
		$_{6}$S$_{31}$ &  8.58831 &     2.65 &     7.56 &  0.02374 & REM \\ 
		\hline
		$_{6}$S$_{32}$ &  8.75932 &     3.13 &     7.27 &  0.01717 & REM \\ 
		\hline
		$_{6}$S$_{33}$ &  8.92692 &     2.62 &     6.87 &  0.01222 & REM \\ 
		\hline
		$_{6}$S$_{34}$ &  9.09208 &     2.92 &     6.34 &  0.00860 & REM \\ 
		\hline
		$_{6}$S$_{35}$ &  9.25799 &     3.61 &     5.70 &  0.00596 & REM \\ 
		\hline
		$_{6}$S$_{36}$ &  9.42384 &     3.10 &     4.93 &  0.00411 & REM \\ 
		\hline
		$_{6}$S$_{37}$ &  9.59800 &     3.50 &     4.06 &  0.00282 & REM \\ 
		\hline
		$_{6}$S$_{38}$ &  9.76062 &     2.89 &     3.08 &  0.00193 & REM \\ 
		\hline
		$_{6}$S$_{39}$ &  9.92885 &     2.27 &     2.02 &  0.00132 & REM \\ 
		\hline
		$_{7}$S$_{5}$ &  3.65753 &     0.02 &     1.95 &  0.53110 & D13 \\ 
		\hline
		$_{7}$S$_{6}$ &  3.95562 &     0.02 &     2.99 &  0.23759 & D13 \\ 
		\hline
		$_{7}$S$_{7}$ &  4.23437 &     0.03 &     3.81 &  0.04431 & D13 \\ 
		\hline
		$_{7}$S$_{8}$ &  4.44941 &     0.13 &     4.05 &  0.21890 & D13 \\ 
		\hline
		$_{7}$S$_{9}$ &  4.61444 &     0.14 &     3.95 &  0.53275 & D13 \\ 
		\hline
		$_{7}$S$_{10}$ &  4.76350 &     4.50 &     3.81 &  0.73377 & REM \\ 
		\hline
		$_{7}$S$_{11}$ &  4.91550 &     3.00 &     3.55 &  0.80700 & REM \\ 
		\hline
		$_{7}$S$_{12}$ &  5.06925 &     1.53 &     3.01 &  0.77668 & REM \\ 
		\hline
		$_{7}$S$_{17}$ &  6.61015 &     3.77 &     3.32 &  1.07422 & REM \\ 
		\hline
	\end{tabular}
	\caption{Continued: properties and composition of the modes in the dataset. Modes are ordered by $n$ and then $l$ with spheroidal modes followed by toroidal modes. Gradients given are for the solid layer. Sources: REM \citep{REM}, M\&W95 \citep{Masters1995}, R\&R98 \citep{Resovsky1998}, D13 \citep{Deuss2013}, K13 \citep{Koelemeijer2013}, S\&D21 \citep{Schneider2021}.}
	\label{dataset_table5}
\end{table}

\pagebreak
\setcounter{table}{0}

\begin{table}[H]
	\centering
	\begin{tabular}{|l|c|c|c|c|c|}
		\hline
		Mode & \makecell{Centre frequency \\ (mHz)} & \makecell{Measurement \\ error, $\epsilon_{d}$ \\(\textmu Hz)} & \makecell{Model error, \\ $\epsilon_{f}$ (\textmu Hz)} & \makecell{Gradient \\ (\textmu Hz/km)} & Source \\ 
		\hline
		$_{7}$S$_{19}$ &  6.91981 &     4.48 &     2.05 &  0.76908 & REM \\ 
		\hline
		$_{7}$S$_{20}$ &  7.07702 &     3.61 &     1.73 &  0.62624 & REM \\ 
		\hline
		$_{7}$S$_{21}$ &  7.24837 &     2.82 &     2.23 &  0.49808 & REM \\ 
		\hline
		$_{7}$S$_{22}$ &  7.41874 &     1.96 &     3.60 &  0.38848 & REM \\ 
		\hline
		$_{7}$S$_{23}$ &  7.59393 &     1.68 &     5.16 &  0.29956 & REM \\ 
		\hline
		$_{7}$S$_{24}$ &  7.77889 &     2.24 &     6.75 &  0.23104 & REM \\ 
		\hline
		$_{7}$S$_{25}$ &  7.96279 &     1.87 &     8.23 &  0.18010 & REM \\ 
		\hline
		$_{7}$S$_{26}$ &  8.15435 &     2.26 &     9.48 &  0.14265 & REM \\ 
		\hline
		$_{7}$S$_{27}$ &  8.34233 &     1.70 &    10.44 &  0.11479 & REM \\ 
		\hline
		$_{7}$S$_{28}$ &  8.52134 &     4.00 &    11.08 &  0.09339 & REM \\ 
		\hline
		$_{7}$S$_{29}$ &  8.70931 &     4.00 &    11.42 &  0.07632 & REM \\ 
		\hline
		$_{7}$S$_{30}$ &  8.90255 &     2.35 &    11.48 &  0.06221 & REM \\ 
		\hline
		$_{7}$S$_{31}$ &  9.08932 &     2.74 &    11.27 &  0.05030 & REM \\ 
		\hline
		$_{7}$S$_{32}$ &  9.27918 &     2.30 &    10.85 &  0.04015 & REM \\ 
		\hline
		$_{7}$S$_{33}$ &  9.45748 &     2.07 &    10.26 &  0.03154 & REM \\ 
		\hline
		$_{7}$S$_{34}$ &  9.63685 &     2.67 &     9.54 &  0.02435 & REM \\ 
		\hline
		$_{7}$S$_{35}$ &  9.82031 &     2.70 &     8.72 &  0.01844 & REM \\ 
		\hline
		$_{8}$S$_{6}$ &  4.43029 &     0.03 &     1.94 &  0.80539 & D13 \\ 
		\hline
		$_{8}$S$_{7}$ &  4.64644 &     0.16 &     0.41 &  1.45558 & D13 \\ 
		\hline
		$_{8}$S$_{10}$ &  5.50302 &     0.04 &     1.62 &  1.28788 & D13 \\ 
		\hline
		$_{8}$S$_{11}$ &  5.70954 &     1.44 &     5.13 &  1.32650 & REM \\ 
		\hline
		$_{8}$S$_{21}$ &  7.97633 &     4.25 &     3.66 &  1.24481 & REM \\ 
		\hline
		$_{8}$S$_{22}$ &  8.12786 &     4.39 &     3.08 &  1.09539 & REM \\ 
		\hline
		$_{9}$S$_{6}$ &  4.61887 &     0.17 &     3.26 &  0.81608 & D13 \\ 
		\hline
		$_{9}$S$_{8}$ &  5.13846 &     0.06 &     7.26 &  0.26303 & D13 \\ 
		\hline
		$_{9}$S$_{9}$ &  5.38154 &     3.00 &     7.07 &  0.22732 & REM \\ 
		\hline
		$_{9}$S$_{10}$ &  5.60608 &     0.23 &     4.97 &  0.08747 & D13 \\ 
		\hline
		$_{9}$S$_{11}$ &  5.88236 &     0.05 &     2.61 &  0.07572 & D13 \\ 
		\hline
		$_{9}$S$_{12}$ &  6.18367 &     0.04 &     2.60 &  0.05106 & D13 \\ 
		\hline
		$_{9}$S$_{13}$ &  6.48069 &     0.05 &     2.42 &  0.00705 & D13 \\ 
		\hline
		$_{9}$S$_{14}$ &  6.76472 &     0.03 &     1.77 &  0.01260 & D13 \\ 
		\hline
		$_{9}$S$_{15}$ &  7.02530 &     0.10 &     1.57 &  0.28308 & D13 \\ 
		\hline
		$_{9}$S$_{16}$ &  7.23273 &     2.51 &     5.14 &  1.02915 & REM \\ 
		\hline
		$_{9}$S$_{18}$ &  7.54147 &     3.86 &     5.30 &  1.52990 & REM \\ 
		\hline
		$_{10}$S$_{8}$ &  5.73500 &     7.00 &     3.46 &  0.95393 & REM \\ 
		\hline
		$_{10}$S$_{9}$ &  5.93899 &     4.50 &     5.45 &  0.97559 & REM \\ 
		\hline
		$_{10}$S$_{10}$ &  6.18646 &     0.21 &     8.12 &  0.67123 & D13 \\ 
		\hline
		$_{10}$S$_{11}$ &  6.44665 &     2.06 &     9.68 &  0.33502 & REM \\ 
		\hline
		$_{10}$S$_{12}$ &  6.68499 &     4.00 &     8.81 &  0.45791 & REM \\ 
		\hline
		$_{10}$S$_{13}$ &  6.86378 &     2.85 &     6.82 &  0.96844 & REM \\ 
		\hline
	\end{tabular}
	\caption{Continued: properties and composition of the modes in the dataset. Modes are ordered by $n$ and then $l$ with spheroidal modes followed by toroidal modes. Gradients given are for the solid layer. Sources: REM \citep{REM}, M\&W95 \citep{Masters1995}, R\&R98 \citep{Resovsky1998}, D13 \citep{Deuss2013}, K13 \citep{Koelemeijer2013}, S\&D21 \citep{Schneider2021}.}
	\label{dataset_table6}
\end{table}

\pagebreak
\setcounter{table}{0}

\begin{table}[H]
	\centering
	\begin{tabular}{|l|c|c|c|c|c|}
		\hline
		Mode & \makecell{Centre frequency \\ (mHz)} & \makecell{Measurement \\ error, $\epsilon_{d}$ \\(\textmu Hz)} & \makecell{Model error, \\ $\epsilon_{f}$ (\textmu Hz)} & \makecell{Gradient \\ (\textmu Hz/km)} & Source \\ 
		\hline
		$_{10}$S$_{15}$ &  7.19800 &     5.00 &     3.72 &  1.37551 & REM \\ 
		\hline
		$_{10}$S$_{16}$ &  7.42013 &     1.52 &     5.79 &  0.77048 & REM \\ 
		\hline
		$_{10}$S$_{17}$ &  7.67267 &     0.07 &     5.88 &  0.41226 & D13 \\ 
		\hline
		$_{10}$S$_{18}$ &  7.93640 &     0.11 &     5.49 &  0.31037 & D13 \\ 
		\hline
		$_{10}$S$_{19}$ &  8.19677 &     0.09 &     4.89 &  0.33182 & D13 \\ 
		\hline
		$_{10}$S$_{20}$ &  8.44607 &     0.16 &     3.84 &  0.49206 & D13 \\ 
		\hline
		$_{10}$S$_{21}$ &  8.67134 &     0.18 &     2.03 &  0.85577 & D13 \\ 
		\hline
		$_{10}$S$_{22}$ &  8.86468 &     2.44 &     2.25 &  1.24961 & REM \\ 
		\hline
		$_{11}$S$_{9}$ &  6.43187 &     0.05 &     2.03 &  1.28565 & D13 \\ 
		\hline
		$_{11}$S$_{10}$ &  6.70557 &     0.11 &     2.03 &  2.52129 & D13 \\ 
		\hline
		$_{11}$S$_{11}$ &  6.91500 &     9.00 &     3.98 &  2.88715 & REM \\ 
		\hline
		$_{11}$S$_{12}$ &  7.14296 &     0.41 &     7.83 &  2.24189 & D13 \\ 
		\hline
		$_{11}$S$_{13}$ &  7.41149 &     4.00 &    11.79 &  1.18150 & REM \\ 
		\hline
		$_{11}$S$_{14}$ &  7.67954 &     0.52 &    13.54 &  0.42144 & D13 \\ 
		\hline
		$_{11}$S$_{17}$ &  8.26500 &     8.00 &     8.28 &  1.34157 & REM \\ 
		\hline
		$_{11}$S$_{20}$ &  8.72419 &     4.78 &     3.89 &  1.65628 & REM \\ 
		\hline
		$_{11}$S$_{21}$ &  8.89697 &     3.35 &     2.81 &  1.28358 & REM \\ 
		\hline
		$_{11}$S$_{22}$ &  9.10307 &     2.35 &     3.60 &  0.83235 & REM \\ 
		\hline
		$_{11}$S$_{23}$ &  9.33287 &     0.25 &     4.27 &  0.58032 & D13 \\ 
		\hline
		$_{11}$S$_{24}$ &  9.57049 &     0.24 &     4.19 &  0.50042 & D13 \\ 
		\hline
		$_{11}$S$_{25}$ &  9.80853 &     0.13 &     3.75 &  0.52349 & D13 \\ 
		\hline
		$_{12}$S$_{6}$ &  5.64385 &     0.15 &     3.37 &  1.23691 & D13 \\ 
		\hline
		$_{12}$S$_{7}$ &  5.85244 &     0.10 &     2.60 &  0.98075 & D13 \\ 
		\hline
		$_{12}$S$_{10}$ &  6.85999 &     5.00 &     5.20 &  0.25527 & REM \\ 
		\hline
		$_{12}$S$_{11}$ &  7.13344 &     0.04 &     5.35 &  0.13445 & D13 \\ 
		\hline
		$_{12}$S$_{12}$ &  7.44891 &     0.04 &     4.39 &  0.44311 & D13 \\ 
		\hline
		$_{12}$S$_{13}$ &  7.76984 &     0.11 &     3.31 &  0.72567 & D13 \\ 
		\hline
		$_{12}$S$_{14}$ &  8.09028 &     0.09 &     2.31 &  0.90544 & D13 \\ 
		\hline
		$_{12}$S$_{15}$ &  8.40452 &     0.09 &     3.22 &  0.59477 & D13 \\ 
		\hline
		$_{12}$S$_{16}$ &  8.68669 &     0.11 &    11.50 &  0.36440 & D13 \\ 
		\hline
		$_{12}$S$_{17}$ &  8.92822 &     0.14 &    13.35 &  0.52073 & D13 \\ 
		\hline
		$_{13}$S$_{15}$ &  8.47266 &     0.63 &    10.65 &  2.21679 & D13 \\ 
		\hline
		$_{13}$S$_{16}$ &  8.74485 &     0.28 &     3.96 &  1.59814 & D13 \\ 
		\hline
		$_{13}$S$_{17}$ &  9.05382 &     1.56 &     3.24 &  0.88172 & REM \\ 
		\hline
		$_{13}$S$_{18}$ &  9.36372 &     0.11 &     3.19 &  0.39762 & D13 \\ 
		\hline
		$_{13}$S$_{19}$ &  9.66449 &     0.14 &     2.98 &  0.13947 & D13 \\ 
		\hline
		$_{13}$S$_{20}$ &  9.95448 &     0.10 &     3.01 &  0.07145 & D13 \\ 
		\hline
		$_{14}$S$_{8}$ &  7.04253 &     0.28 &     4.33 &  0.58137 & D13 \\ 
		\hline
		$_{14}$S$_{9}$ &  7.34452 &     0.12 &     5.65 &  0.35213 & D13 \\ 
		\hline
		$_{14}$S$_{13}$ &  8.72982 &     0.07 &     7.88 &  0.10955 & D13 \\ 
		\hline
	\end{tabular}
	\caption{Continued: properties and composition of the modes in the dataset. Modes are ordered by $n$ and then $l$ with spheroidal modes followed by toroidal modes. Gradients given are for the solid layer. Sources: REM \citep{REM}, M\&W95 \citep{Masters1995}, R\&R98 \citep{Resovsky1998}, D13 \citep{Deuss2013}, K13 \citep{Koelemeijer2013}, S\&D21 \citep{Schneider2021}.}
	\label{dataset_table7}
\end{table}

\pagebreak
\setcounter{table}{0}

\begin{table}[H]
	\centering
	\begin{tabular}{|l|c|c|c|c|c|}
		\hline
		Mode & \makecell{Centre frequency \\ (mHz)} & \makecell{Measurement \\ error, $\epsilon_{d}$ \\(\textmu Hz)} & \makecell{Model error, \\ $\epsilon_{f}$ (\textmu Hz)} & \makecell{Gradient \\ (\textmu Hz/km)} & Source \\ 
		\hline
		$_{14}$S$_{14}$ &  8.98149 &     0.08 &     7.93 &  0.23474 & D13 \\ 
		\hline
		$_{15}$S$_{11}$ &  8.12241 &     1.50 &     6.10 &  1.35115 & REM \\ 
		\hline
		$_{15}$S$_{12}$ &  8.42773 &     0.13 &     7.15 &  0.63755 & D13 \\ 
		\hline
		$_{15}$S$_{15}$ &  9.59215 &     0.11 &     3.39 &  3.22890 & D13 \\ 
		\hline
		$_{15}$S$_{16}$ &  9.92112 &     0.16 &     2.16 &  3.16762 & D13 \\ 
		\hline
		$_{16}$S$_{10}$ &  8.43336 &     0.08 &     5.47 &  0.35160 & D13 \\ 
		\hline
		$_{16}$S$_{11}$ &  8.73013 &     0.27 &     5.76 &  1.04709 & D13 \\ 
		\hline
		$_{16}$S$_{14}$ &  9.29932 &     0.95 &     3.28 &  3.80556 & D13 \\ 
		\hline
		$_{17}$S$_{12}$ &  9.14844 &     0.06 &     7.25 &  0.67647 & D13 \\ 
		\hline
		$_{17}$S$_{13}$ &  9.42847 &     0.08 &     9.13 &  0.47829 & D13 \\ 
		\hline
		$_{17}$S$_{14}$ &  9.69854 &     0.30 &     9.30 &  0.60608 & D13 \\ 
		\hline
		$_{17}$S$_{15}$ &  9.93267 &     0.52 &     8.21 &  0.93383 & D13 \\ 
		\hline
		$_{18}$S$_{9}$ &  8.73500 &     7.00 &     5.91 &  1.66112 & REM \\ 
		\hline
		$_{19}$S$_{11}$ &  9.64479 &     0.23 &     5.65 &  0.16856 & D13 \\ 
		\hline
		$_{1}$T$_{2}$ &  1.31929 &     0.17 &     0.46 &  0.17728 & S\&D21 \\ 
		\hline
		$_{1}$T$_{3}$ &  1.43845 &     0.08 &     0.48 &  0.22153 & S\&D21 \\ 
		\hline
		$_{1}$T$_{5}$ &  1.75019 &     0.13 &     0.57 &  0.26744 & S\&D21 \\ 
		\hline
		$_{1}$T$_{6}$ &  1.92525 &     0.12 &     0.62 &  0.25168 & S\&D21 \\ 
		\hline
		$_{1}$T$_{7}$ &  2.10340 &     0.06 &     0.67 &  0.21273 & S\&D21 \\ 
		\hline
		$_{1}$T$_{13}$ &  3.09907 &     0.10 &     0.86 &  0.01422 & S\&D21 \\ 
		\hline
		$_{1}$T$_{14}$ &  3.25352 &     0.06 &     0.82 &  0.00776 & S\&D21 \\ 
		\hline
		$_{2}$T$_{3}$ &  2.29528 &     0.22 &     0.33 &  0.35163 & S\&D21 \\ 
		\hline
		$_{2}$T$_{5}$ &  2.48487 &     0.15 &     0.69 &  0.49077 & S\&D21 \\ 
		\hline
		$_{2}$T$_{7}$ &  2.75300 &     0.33 &     1.14 &  0.64240 & S\&D21 \\ 
		\hline
		$_{2}$T$_{8}$ &  2.91257 &     0.08 &     1.36 &  0.68896 & S\&D21 \\ 
		\hline
		$_{2}$T$_{13}$ &  3.83052 &     0.21 &     1.69 &  0.35745 & S\&D21 \\ 
		\hline
		$_{2}$T$_{14}$ &  4.01381 &     0.24 &     1.56 &  0.25537 & S\&D21 \\ 
		\hline
		$_{2}$T$_{15}$ &  4.19269 &     0.22 &     1.37 &  0.17272 & S\&D21 \\ 
		\hline
		$_{2}$T$_{16}$ &  4.36875 &     0.53 &     1.13 &  0.11167 & S\&D21 \\ 
		\hline
		$_{3}$T$_{1}$ &  3.20008 &     1.73 &     1.86 &  0.46639 & S\&D21 \\ 
		\hline
		$_{3}$T$_{7}$ &  3.60368 &     0.24 &     1.84 &  0.84837 & S\&D21 \\ 
		\hline
		$_{3}$T$_{16}$ &  5.05139 &     0.23 &     0.65 &  1.00943 & S\&D21 \\ 
		\hline
		$_{4}$T$_{9}$ &  4.77537 &     1.86 &     0.11 &  1.10294 & S\&D21 \\ 
		\hline
	\end{tabular}
	\caption{Continued: properties and composition of the modes in the dataset. Modes are ordered by $n$ and then $l$ with spheroidal modes followed by toroidal modes. Gradients given are for the solid layer. Sources: REM \citep{REM}, M\&W95 \citep{Masters1995}, R\&R98 \citep{Resovsky1998}, D13 \citep{Deuss2013}, K13 \citep{Koelemeijer2013}, S\&D21 \citep{Schneider2021}.}
	\label{dataset_table8}
\end{table}

\pagebreak
\bibliography{bibliography}